\documentclass{naturebsg}
\pdfoutput=1

\usepackage{amsmath}
\usepackage{amssymb}
\usepackage[normalem]{ulem}
\usepackage{color}
\usepackage{times}
\usepackage{pifont}
\usepackage{graphics}
\usepackage[svgnames]{xcolor}
\usepackage{epsfig}
\usepackage{threeparttable}

\newcommand{\mstar}{\ensuremath{\,{M}_{*}}}
\newcommand{\rstar}{\ensuremath{\,{R}_{*}}}

\newcommand{\teff}{\ensuremath{T_{\rm eff}}}
\newcommand{\fbol}{\ensuremath{F_{\rm bol}}}
\newcommand{\teq}{\ensuremath{T_{\rm eq}}}

\newcommand{\loggstar}{\ensuremath{\log{g_\star}}}
\newcommand{\msun}{\ensuremath{\,{M}_\odot}}
\newcommand{\rsun}{\ensuremath{\,{R}_\odot}}
\newcommand{\lsun}{\ensuremath{\,{L}_\odot}}
\newcommand{\vsini}{\ensuremath{\,{v}\sin{I_*}}}

\newcommand{\feh}{\ensuremath{\left[{\rm Fe}/{\rm H}\right]}}

\newcommand{\mj}{\ensuremath{\,M_{\rm J}}}
\newcommand{\rj}{\ensuremath{\,R_{\rm J}}}
\newcommand{\fave}{\langle F \rangle}
\newcommand{\fluxcgs}{10$^9$ erg s$^{-1}$ cm$^{-2}$}
\newcommand{\mhosta}{2.5}
\newcommand{\rhosta}{2.4}

\newcommand{\teffa}{10,170}

\newcommand{\rplaneta}{1.9}
\newcommand{\teqplaneta}{4050}
\newcommand{\perioda}{1.48}

\newcommand\arcmin{\mbox{$^\prime$}}%
\newcommand\farcs{\mbox{$.\!\!^{\prime\prime}$}}%
\newcommand\fs{\mbox{$.\!\!^{\mathrm s}$}}%
\newcommand{\apj}{\it The Astrophysical Journal}
\newcommand{\apjl}{\it The Astrophysical Journal Letters}
\newcommand{\mnras}{\it Monthly Notices of the Royal Astronomical Society}
\newcommand{\aap}{\it Astronomy \& Astrophysics}

\newcommand{\apss}{\it Astronomy \& Astrophysics Supplements}
\newcommand{\apjs}{\it The Astrophysical Journal Supplements}
\newcommand{\pasp}{\it Publications of the Astronomical Society of the Pacific}

\newcommand{\degr}{\ensuremath{^\circ}}

\title{\spacing{1}{\Large\bfseries\noindent\sloppy\textsf{A giant planet undergoing extreme ultraviolet irradiation by its hot massive-star host}}}

\author{}
\date{}

\begin{document}

\def\@cite#1#2{$^{\mbox{\scriptsize #1\if@tempswa , #2\fi}}$}


\maketitle

{\noindent\sloppy
B. Scott Gaudi$^{1}$, 
Keivan G. Stassun$^{2,3}$, 
Karen A. Collins$^{2}$, 
Thomas G. Beatty$^{4,5}$,
George Zhou$^{6}$,
David W. Latham$^{6}$,
Allyson Bieryla$^{6}$,
Jason D. Eastman$^{6}$,
Robert J. Siverd$^{7}$,
Justin R. Crepp$^{8}$,
Erica J. Gonzales$^{8}$,
Daniel J. Stevens$^{1}$,
Lars A. Buchhave$^{9,10}$,
Joshua Pepper$^{11}$,
Marshall C. Johnson$^{1}$,
Knicole D. Colon$^{12,13}$,
Eric L. N. Jensen$^{14}$,
Joseph E. Rodriguez$^{6}$,
Valerio Bozza$^{15,16}$,
Sebastiano Calchi Novati$^{15,17}$,
Giuseppe D`Ago$^{18,19}$,
Mary T. Dumont$^{20,21}$,
Tyler Ellis$^{22,23}$,
Clement Gaillard$^{20}$,
Hannah Jang-Condell$^{22}$,
David H. Kasper$^{22}$,
Akihiko Fukui$^{24}$,
Joao Gregorio$^{25}$,
Ayaka Ito$^{26,27}$,
John F. Kielkopf$^{28}$,
Mark Manner$^{29}$,
Kyle Matt$^{20}$,
Norio Narita$^{26,30,31}$,
Thomas E. Oberst$^{32}$,
Phillip A. Reed$^{33}$,
Gaetano Scarpetta$^{15,17}$,
Denice C. Stephens$^{20}$,
Rex R. Yeigh$^{22}$,
Roberto Zambelli$^{34}$,
B.J. Fulton$^{35,36}$,
Andrew W. Howard$^{35}$,
David J. James$^{37}$,
Matthew Penny$^{1,38}$, 
Daniel Bayliss$^{39}$, 
Ivan A. Curtis$^{40}$, 
D.L. DePoy$^{41}$,
Gilbert A. Esquerdo$^{6}$,
Andrew Gould$^{1,42}$,
Michael D. Joner$^{20}$,
Rudolf B. Kuhn$^{43}$,
Jonathan Labadie-Bartz$^{11}$,
Michael B. Lund$^{2}$,
Jennifer L. Marshall$^{41}$,
Kim K. McLeod$^{44}$,
Richard W. Pogge$^{1}$,
Howard Relles$^{6}$,
Chistopher Stockdale$^{45}$,
T.G. Tan$^{46}$,
Mark Trueblood$^{47}$,
Patricia Trueblood$^{47}$
}

\bigskip
\renewcommand{\baselinestretch}{1.0}
\begin{affiliations}
\item Department of Astronomy, The Ohio State University, Columbus, OH, 43210, USA 
\item Department of Physics and Astronomy, Vanderbilt University, 6301 Stevenson Center, Nashville, TN 37235, USA 
\item Department of Physics, Fisk University, 1000 17th Avenue North, Nashville, TN 37208, USA
\item Center for Exoplanets and Habitable Worlds, The Pennsylvania State University, 525 Davey Lab, University Park, PA 16802, USA
\item Department of Astronomy \& Astrophysics, The Pennsylvania
State University, 525 Davey Lab, University Park, PA 16802, USA
\item Harvard-Smithsonian Center for Astrophysics, 60 Garden Street, Cambridge, MA 02138, USA
\item Las Cumbres Observatory Global Telescope Network, 6740 Cortona Dr., Suite 102, Santa Barbara, CA 93117, USA
\item Department of Physics, University of Notre Dame, 225 Nieuwland Science Hall, Notre Dame, IN 46556, USA
\item Niels Bohr Institute, University of Copenhagen, Juliane Maries vej 30, 21S00 Copenhagen, Denmark
\item Centre for Star and Planet Formation, Geological Museum, Øster Voldgade 5, 1350 Copenhagen, Denmark
\item Department of Physics, Lehigh University, 16 Memorial Drive East, Bethlehem, PA 18015, USA
\item NASA Ames Research Center, M/S 244-30, Moffett Field, CA 94035, USA
\item Bay Area Environmental Research Institute, 625 2nd St. Ste 209 Petaluma, CA 94952, USA
\item Department of Physics and Astronomy, Swarthmore College, Swarthmore, PA 19081, USA
\item Dipartimento di Fisica ``E. R. Caianiello'', Universit\`a di Salerno, Via Giovanni Paolo II 132, 84084 Fisciano (SA), Italy
\item Istituto Nazionale di Fisica Nucleare, Sezione di Napoli, 80126 Napoli, Italy
\item IPAC, Mail Code 100-22, Caltech, 1200 E. California Blvd., Pasadena, CA 91125
\item Istituto Internazionale per gli Alti Studi Scientifici (IIASS), Via G. Pellegrino 19, 84019 Vietri sul Mare (SA), Italy
\item INAF-Observatory of Capodimonte, Salita Moiariello, 16, 80131, Naples, Italy
\item Department of Physics and Astronomy, Brigham Young University, Provo, UT 84602, USA
\item Department of Astronomy and Astrophysics, University of California Santa Cruz, Santa Cruz, CA, 95064, USA
\item Department of Physics and Astronomy, University of Wyoming, 1000 E. University, Laramie, WY 82071, USA
\item Department of Physics \& Astronomy, Louisiana state University, 202 Nicholson Hall, Baton Rouge, LA 70803, USA
\item Okayama Astrophysical Observatory, National Astronomical Observatory of Japan, NINS,
Asakuchi, Okayama 719-0232, Japan
\item Atalaia Group \& Crow-Observatory, Portalegre, Portugal
\item National Astronomical Observatory of Japan, NINS, 2-21-1 Osawa, Mitaka, Tokyo 181-8588, Kanto, Japan
\item Graduate School of Science and Engineering, Hosei University, 3-7-2 Kajino-cho, Koganeishi, Tokyo 184-8584, Japan
\item Department of Physics and Astronomy, University of Louisville, Louisville, KY 40292, USA
\item Spot Observatory, Nashville, TN 37206 USA
\item Department of Astronomy, The University of Tokyo, 7-3-1 Hongo, Bunkyo-ku, Tokyo 113-0033, Japan
\item Astrobiology Center, NINS, 2-21-1 Osawa, Mitaka, Tokyo 181-8588, Japan
\item Department of Physics, Westminster College, New Wilmington, PA, 16172, USA
\item Department of Physical Sciences, Kutztown University, Kutztown, PA 19530, USA
\item Societ{\`a} Astronomica Lunae, Castelnuovo Magra 19030, Italy
\item Institute for Astronomy, University of Hawaii, 2680 Woodlawn Drive, Honolulu, HI 96822-1839, USA
\item NSF Graduate Research Fellow
\item Astronomy Department, University of Washington, Box 351580, Seattle, WA 98195, USA
\item Sagan Fellow
\item Observatoire Astronomique de l'Universit{\'e} de Gen{\`e}ve, 51 Chemin des Maillettes, 1290 Versoix, Switzerland
\item ICO, Adelaide, Australia
\item George P. and Cynthia Woods Mitchell Institute for Fundamental Physics and Astronomy, and Department of Physics and Astronomy, Texas A \& M University, College Station, TX 77843-4242, USA
\item Max Planck Institute for Astronomy, K{\"o}nigstuhl 17, D-69117 Heidelberg, Germany
\item South African Astronomical Observatory, PO Box 9, Observatory 7935, South Africa
\item Wellesley College, 106 Central St, Wellesley, MA 02481, USA
\item Hazelwood Observatory, Victoria, Australia
\item Perth Exoplanet Survey Telescope, Perth, Australia
\item Winer Observatory, Sonoita, AZ 85637, USA
\end{affiliations}
\renewcommand{\baselinestretch}{1.0}

\begin{abstract}
The amount of ultraviolet irradiation and ablation experienced by a planet depends strongly on the temperature of its host star. Of the thousands of extra-solar planets now known, only four giant planets have been found that transit hot, A-type stars (temperatures of 7300--10,000~K), and none are known to transit even hotter B-type stars. 
WASP-33 is an A-type star with a temperature of $\sim$7430~K, which hosts the hottest known transiting planet\cite{Collier:2010a}; the planet is itself as hot as a red dwarf star of type M\cite{Vonessen:2015}. The planet displays a large heat differential between its day-side and night-side\cite{Vonessen:2015}, and is highly inflated, traits that have been linked to high insolation\cite{Komacek:2016,Demory:2011}. However, even at the temperature of WASP-33b's day-side, its atmosphere likely resembles the molecule-dominated atmospheres of other planets, and at the level of ultraviolet irradiation it experiences, its atmosphere is unlikely to be significantly ablated over the lifetime of its star. Here we report observations of the bright star HD~195689, which reveal a close-in (orbital period $\sim$\perioda~days) transiting giant planet, KELT-9b. At $\sim$\teffa~K, the host star is at the dividing line between stars of type A and B, and we measure the KELT-9b's day-side temperature to be $\sim$4600~K. This is as hot as stars of stellar type K4\cite{Pecaut:2013}. The molecules in K stars are entirely dissociated, and thus the primary sources of opacity in the day-side atmosphere of KELT-9b are likely atomic metals. Furthermore, KELT-9b receives $\sim$700 times more extreme ultraviolet radiation (wavelengths shorter than 91.2 nanometers) than WASP-33b, leading to a predicted range of mass-loss rates that could leave the planet largely stripped of its envelope during the main-sequence lifetime of the host star\cite{Murray-Clay:2009}.
\end{abstract}

The first transiting planets were discovered around cool, solar-type stars\cite{Charbonneau:2000,Henry:2000}, primarily because hot stars have few spectral lines and rotate rapidly, making Doppler confirmation of planets more difficult. 
Only in the past few years have transiting planets been confirmed around hot stars of types early-F and A\cite{Hartman:2015,Zhou:2017}, inspired by the discovery of WASP-33b\cite{Collier:2010a}. That discovery demonstrated that it is possible to confirm transiting planets around rapidly rotating hot stars via a combination of relatively low-precision radial-velocity measurements
and Doppler tomography.
However, even the hottest of these few A-type transiting-planet host stars only reach temperatures of $\sim$7500~K. Thus, 
while transit surveys, in particular {\it Kepler}\cite{Borucki:2010}, have extended the census of planets around low-mass stars, our understanding of planets around massive, hot stars remains poor. 

Massive stars cool and spin down as they evolve, enabling precise Doppler measurements. Thus the
primary strategy to search for planets around high-mass stars has been  surveys of 
``retired A-stars"\cite{Johnson:2007}, high-mass stars that have already evolved into subgiant and giant stars. These stars have revealed a paucity of short-period giant planets relative to sun-like main-sequence stars\cite{Johnson:2007}.  One interpretation is that the initial planet population of high-mass stars is similar to that seen in unevolved sun-like stars, but that the short-period planets are subsequently engulfed during the evolution of their parent stars 
or ablated by the intense irradiation of their host stars while they are still hot\cite{Murray-Clay:2009}. Another interpretation is that these stars actually have
masses similar to the Sun\cite{Lloyd:2011},  
implying that the paucity of short-period planets among the retired A-stars is indeed a signature of planet engulfment\cite{Schlaufman:2013}, as sun-like stars
do not emit strong ultraviolet radiation with which to ablate their planets\cite{Murray-Clay:2009}. 

It is therefore critical to assay the population of short-period planets around {\it bona fide}
high-mass stars while they are still on the main sequence,
and then to map the evolution of these planets through to their later evolutionary phases.
Although there have been radial-velocity surveys targeting unevolved high-mass stars\cite{Galland:2005,Borgniet:2017}, there are still no known transiting planets around unevolved stars more massive than 2~\msun\ that produce
high levels of extreme ultraviolet irradiation.

\begin{figure}
\includegraphics[width=\linewidth]{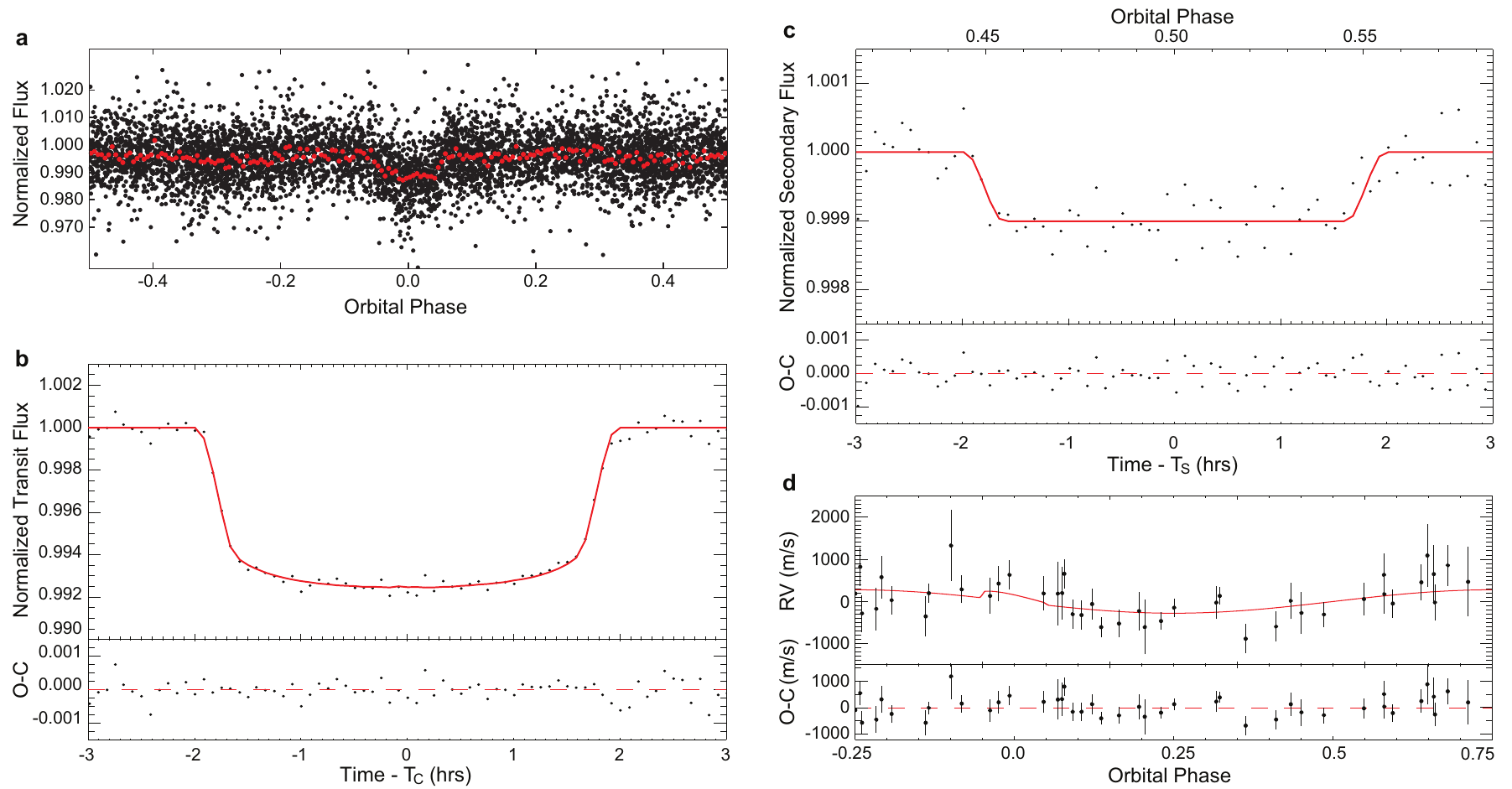}
\caption{
{\bf KELT-9 discovery and confirmation data.} Shown are the discovery light curve, follow-up primary transit and secondary eclipse light curves, and reflex Doppler signal due to the transiting companion.
(a) Field 11 of the northern KELT survey was observed 6001 times from UT 2007 May 30 to UT 2013 June 14, following the same procedures as described for the first KELT planet\cite{Siverd:2012}  The black points show the unbinned data folded at the period of the planet (P = 1.4810932 days), while the red points show these points binned in phase using a bin size of 0.05 in phase. 
(Methods).
(b) Combined, binned follow-up light curves of the primary transit, where TC is the time of the center of the transit. The transit shape is consistent with an opaque, circular planet occulting a star with the limb darkening expected for KELT-9's spectral type of B9.5--A0 The black points show the average of all follow-up light curves, combined in 5-minute bins. The combined best-fit models binned the same way are shown as a solid red line.
(c) Combined, binned follow-up light curves of the secondary eclipse.  The magnitude of the signal gives a temperature for the day-side of the planet of $\sim$4600~K. This is consistent with the range of expected equilibrium temperatures of the planet and suggests poor heat redistribution to the night side of the planet. The data points and red curve are binned in the same manner as in b. TS is the time of the center of the secondary eclipse.
(d) Doppler reflex curve of the primary star. Only the data that were not used in the Doppler Tomography analysis (see Fig.~\ref{fig:dt}) are shown and were used to measure the Doppler reflex curve. The best-fit model of the Doppler reflex curve is shown as the red line.  Note the weak Rossiter-McLaughlin signal near phase zero, which is a prediction based on the Doppler Tomographic signal exhibited in Figure~\ref{fig:dt}. A periodogram of this data alone yields a significant peak with an emphemeris that matches that of the photometric transit data to within $\lesssim 10^{-4}$ days, thereby confirming the reflex signal is due to the transiting planet and verifying the planet mass estimate.
\label{fig:data}
}
\end{figure}

\begin{figure}
\centering
\includegraphics[width=\linewidth]{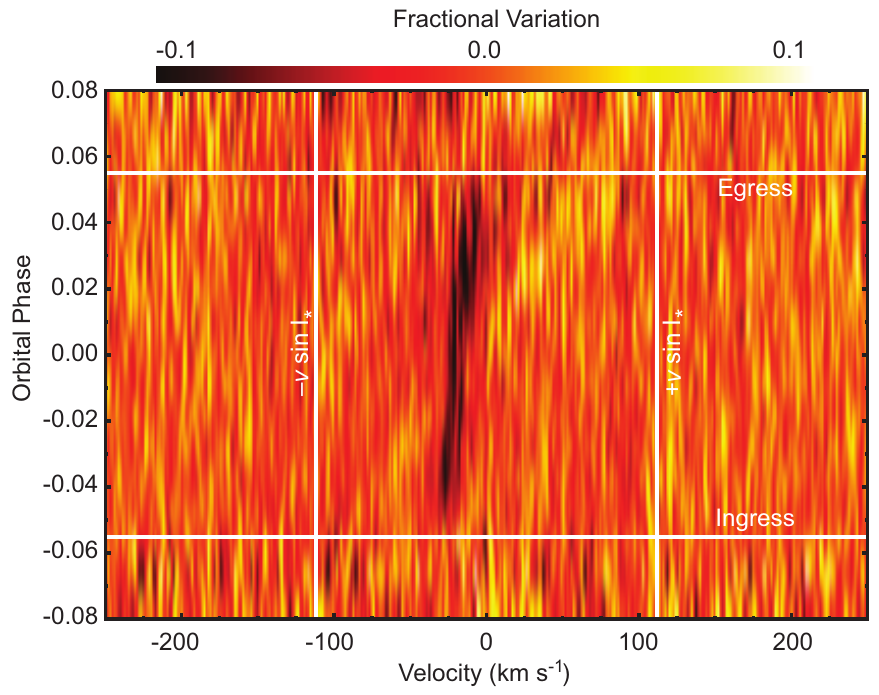}
\caption{
{\bf Combined Doppler tomographic measurements from three separate transits.}  All three transits (see Methods) clearly show a ``Doppler shadow" at the time of transit, with the velocity width as expected given the spectroscopically-measured \vsini\ and the photometrically-measured transit depth and impact parameter, thus confirming the reality of the Doppler tomographic signal and that the planet orbits KELT-9. The color coding shows the fractional variation in the spectroscopic signal from the null hypothesis of no shadow due to a transiting planet.  Darker regions indicate the Doppler shadow as the planet crosses the face of the host star.  The horizontal line indicate the expected beginning (bottom) and end (top) of the transit, and the vertical lines indicate the negative and positive width of the spectral lines due to broadening caused by the project rotation speed of the star (see Methods).  The projected path of the planet across the face of the star is nearly coincident with the stellar rotation axis, with an impact parameter in units of the stellar radius of $\sim 0.2$ and a projected spin-orbit misalignment of $\sim -85$ degrees, indicating that the planet is on a nearly polar orbit. The host star's rapid rotation and resulting oblateness implies that the orbit of the planet is likely to exhibit precession, which should be detectable by the year 2022\cite{Johnson:2015}.
\label{fig:dt}
}
\end{figure}

The Kilodegree Extremely Little Telescope (KELT) is an
all-sky survey for planets transiting bright (visual
magnitude 8--11) stars\cite{Pepper:2007,Pepper:2012}.  
HD~195689 (hereafter KELT-9),  
exhibited repeating transit-like events of $\sim$0.6\% depth with a period of $P\sim$\perioda~d (Figure~\ref{fig:data}), and was selected
as a candidate transiting planet (see the Methods).
KELT-9's basic properties (Table~\ref{tab:k9prop}) include a
high effective temperature and rapid rotation.  Following the approach that led to the discovery of WASP-33b,
we obtained follow-up observations (see Figure~\ref{fig:data}, Figure~\ref{fig:dt}, and the Methods) 
that ultimately confirmed KELT-9b as a transiting planet. 

\renewcommand{\baselinestretch}{1.0}
\renewcommand{\baselinestretch}{1.0}

\begin{table}
\footnotesize
\centering
\begin{tabular}{llr}
\hline
Stellar Parameters: & & \\
                               ~~~$M_{*}$\dotfill &Mass (\msun)\dotfill & $2.52_{-0.20}^{+0.25}$\\
                             ~~~$R_{*}$\dotfill &Radius (\rsun)\dotfill & $2.362_{-0.063}^{+0.075}$\\
                         ~~~$L_{*}$\dotfill &Luminosity (\lsun)\dotfill & $53_{-10}^{+13}$\\
                             ~~~$\rho_*$\dotfill &Density (cgs)\dotfill & $0.2702\pm0.0029$\\
                  ~~~$\log{g_*}$\dotfill &Surface gravity (cgs)\dotfill & $4.093\pm0.014$\\
                  ~~~$\teff$\dotfill &Effective temperature (K)\dotfill & $10170\pm450$\\
                                 ~~~$\feh$\dotfill &Metallicity\dotfill & $-0.03\pm0.20$\\
             ~~~$v\sin{I_*}$\dotfill &Rotational velocity (km/s)\dotfill & $111.4\pm1.3$\\
           ~~~$\lambda$\dotfill &Spin-orbit alignment (degrees)\dotfill & $-84.8\pm1.4$\\
Planetary Parameters: & & \\
                                  ~~~$P$\dotfill &Period (days)\dotfill & $1.4811235\pm0.0000011$\\
                           ~~~$a$\dotfill &Semi-major axis (AU)\dotfill & $0.03462_{-0.00093}^{+0.00110}$\\
                                 ~~~$M_{P}$\dotfill &Mass (\mj)\dotfill & $2.88\pm0.84$\\
                               ~~~$R_{P}$\dotfill &Radius (\rj)\dotfill & $1.891_{-0.053}^{+0.061}$\\
                           ~~~$\rho_{P}$\dotfill &Density (cgs)\dotfill & $0.53\pm0.15$\\
                      ~~~$\log{g_{P}}$\dotfill &Surface gravity\dotfill & $3.30_{-0.15}^{+0.11}$\\
               ~~~$T_{eq}$\dotfill &Equilibrium temperature (K)\dotfill & $4050\pm180$\\
                   ~~~$\fave$\dotfill &Incident flux (\fluxcgs)\dotfill & $61.1_{-9.8}^{+11.0}$\\
\hline
\end{tabular}
\caption{\label{tab:k9prop}
\scriptsize{\sffamily
{\bf Median values and 68\% confidence intervals for the physical properties of the KELT-9 system}  
from a global fit to the light curves, radial velocities, and Doppler tomography data, using a 
custom version of the EXOFAST transit fitting code\cite{Eastman:2013} 
(Methods). 
We use the Yonsei-Yale (YY) isochrones\cite{Demarque:2004} as well as empirically-calibrated stellar relations\cite{Torres:2010} as constraints. In addition, KELT-9 has a measured parallax from {\it Hipparcos}\cite{vanLeeuwen:2007} and {\it Gaia}\cite{Gaia:2016}, thereby allowing us to estimate the stellar radius empirically\cite{Stassun:2016} from its distance, effective temperature, bolometric (wavelength-integrated) flux, and interstellar extinction (Methods). 
We adopt the values derived using constraints from the YY isochrones and Hipparcos-derived radius as our fiducial system parameters.}
}
\normalsize
\end{table}

\begin{figure}
\centering
\includegraphics[width=\linewidth]{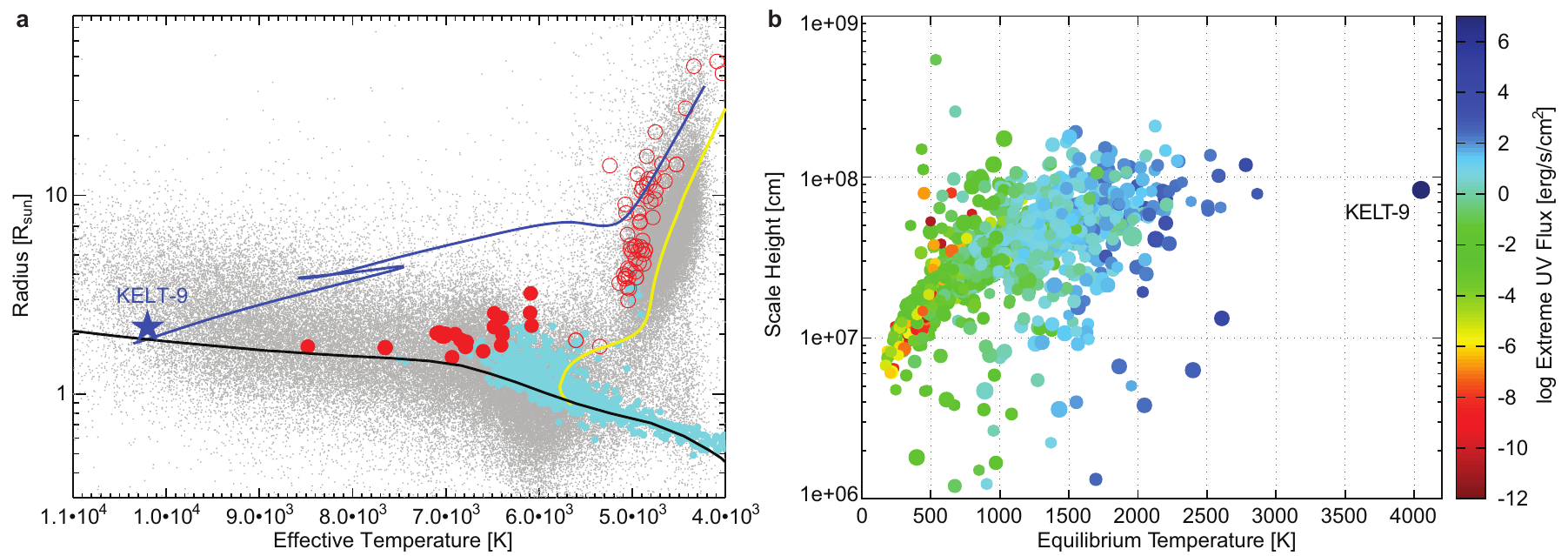}
\caption{{\bf The extreme properties of the KELT-9 system in context and the prospects for follow-up.}
(a) Hertzsprung-Russell (\rstar\ vs.\ \teff) diagram of hosts of known planets detected by the radial-velocity (open circles) and transit methods (filled circles), as well as nearby stars in the Hipparcos catalog (grey points).  Only planet-hosting systems with $V\le 10$ are shown for clarity.
Cyan symbols are low-mass planet hosts with $M < 1.5~M_\odot$, red symbols indicate massive planet hosts with $M_* \ge 1.5~M_\odot$. 
The yellow line shows the evolutionary trajectory for a solar analog ($M_*=M_\odot$ and solar metallicity), whereas the blue track shows the evolutionary trajectory for an analog of KELT-9 ($M_*=\mhosta$~$M_\odot$ and solar metallicity).
We also show the zero-age main sequence for solar-metallicity stars from the YY isochrones (black curve).  KELT-9 is hotter than any other known transiting planet host by $\sim 1500$~K. 
(b) Atmospheric scale height versus equilibrium temperature for known transiting planets with measured masses. 
Color represents the amount of incident extreme ultraviolet ($\lambda \le 91.2$~nanometers) flux the planet receives from its parent star, and the symbol size is inversely proportional to the $V$ magnitude of the host. KELT-9b is hotter than any other known transiting gas giant planet by $\sim 1000$~K and receives $\sim 700$ times more extreme ultraviolet flux. 
\label{fig:properties}
}
\end{figure}

KELT-9 is a hot ($\teff\simeq\teffa$~K), massive ($\mstar \simeq \mhosta~\msun$) star of spectral type B9.5--A0, with a relatively young age of 300 million years (Methods), 
comfortably in its 500-million-year main-sequence phase of evolution;
it has evidently not yet begun its evolution toward becoming a ``retired A star".
Indeed, from comparison with other known planet-hosting stars (Fig.~\ref{fig:properties}),
it is clear that KELT-9 is a likely progenitor of at least a subset of the putative ``retired A star" 
hosts of planets detected by radial-velocity surveys. KELT-9, and other A-star transiting planet hosts, thereby provides an important missing link between these samples of planets, and planets detected in more traditional radial-velocity surveys of sun-like stars.

KELT-9 is only the fifth A-type star 
known to host a transiting giant companion, and is by a significant margin the hottest, most massive, and most luminous known transiting giant planet host. The host star also has the brightest $V$-band magnitude of any transiting hot Jupiter host, being slightly brighter in $V$ than HD~209458b\cite{Charbonneau:2000,Henry:2000}. 

Given the high stellar luminosity and close orbit, the planet receives a large stellar insolation flux (Table~\ref{tab:k9prop}). As a result, it has an extremely high equilibrium temperature, assuming zero albedo and perfect heat redistribution, of $\sim$\teqplaneta~K. 
This is as hot as a late K-type star\cite{Pecaut:2013}, and thus we expected a large thermal emission signal, which we easily confirmed with our $z'$-band detection of the secondary eclipse with a depth of $\sim 0.1\%$ (Figure~\ref{fig:data}). This measurement implies an even hotter day-side temperature 
of $\sim 4600 \pm 150$~K, likely indicating poor 
redistribution of energy to the night side of the planet and a temperature closer to that of a mid-K star. The planet is also extremely inflated relative to theoretical models, with a radius of $\sim \rplaneta~R_{\rm J}$.  Poor redistribution of heat and radius inflation (both of which are also observed in WASP-33b\cite{Collier:2010a,Vonessen:2015}), have been linked to high stellar insolation\cite{Komacek:2016,Demory:2011}, although the exact physical mechanisms remain uncertain.

Thus, although other transiting planets have been found around A-type stars, and indeed some (such as WASP-33b) have been discovered with the temperatures of low-mass stars, 
the KELT-9 planet and host star are hotter by $\sim 1000$~K and $\sim 2500$~K, respectively, than any other known transiting gas-giant system. Consequently, one expects all of the opacity sources on the day-side to be atomic, as in a K-type star. In contrast, all other known transiting planets, which have day-side temperatures of $<3850$~K\cite{Pecaut:2013}, are cool enough to contain molecular species. Furthermore, the very high flux of extreme ultraviolet radiation (wavelengths shorter than 91.2 nanometers) from KELT-9, $\sim 700$ times higher than WASP-33, may lead to unique photochemistry in the planet atmosphere\cite{Casewell:2015}. This places KELT-9b in a qualitatively new regime of planetary atmospheres, and makes characterization of the atmosphere of KELT-9b particularly compelling.  

Fortunately, the brightness of KELT-9 and the extreme properties of its transiting planet make the prospects for detailed characterization of this system promising.  Observations using ground-based facilities, {\it Spitzer}, the {\it Hubble Space Telescope (HST)}, and ultimately the {\it James Webb Space Telescope}, will allow for the measurement of the phase-resolved spectrum of its thermal emission from the far-optical through the infrared ($\sim$30~$\mu$m). 
The low surface gravity of KELT-9b combined with the high temperature lead to one of the largest atmospheric scale heights of any known transiting planet (Figure~\ref{fig:properties}).  The expected few-percent variations in the transmission spectrum during the primary transit should be easily detectable. 

The future evolution of the KELT-9 system is uncertain but certainly interesting. The high ultraviolet flux impinging on KELT-9b likely means that its atmosphere is being significantly ablated, with an estimated mass loss rate\cite{Murray-Clay:2009} 
of $\sim 10^{10}$--$10^{13}~{\rm g~s^{-1}}$. At the upper end of these rates, the planet may be completely stripped of its outer envelope in $<600$~Myr, roughly the time scale for the host to evolve from the main-sequence to the base of the red giant branch (see Methods). These estimates are so uncertain because the physics of planet evaporation is extraordinarily complex, particularly due to the unknown magnitude of the host star activity, and thus very high-energy (extreme ultraviolet and X-ray) non-thermal radiation and stellar wind. In any event, even the lower end of mass-loss rates we estimate should be easily detectable and measurable with HST observations.

On the other hand, as KELT-9 eventually exhausts its core hydrogen supply in $\sim$200 million years, it will grow from its current radius of $\sim$\rhosta~$R_\odot$ to a radius of $\sim 5~R_\odot$, while simultaneously cooling to $\teff\sim 8000$~K. 
Soon after, it will rapidly traverse the `subgiant branch' whereby it will cool to $\sim 5000$~K and expand to $\sim 8~R_\odot$. As it reaches the base of the red giant branch, the stellar surface of KELT-9 will encroach upon the orbit of KELT-9b. Exactly what will happen to the star and planet at this point is far from clear. If the mass loss from ablation is lower than estimated above, the planet may remain intact as a gas giant, and it will be engulfed by its host star, perhaps leading to a bright transient event\cite{Metzger:2012}, and an 
unusually rapidly rotating red giant star with enriched lithium provided by the dissolved planetary companion\cite{Anguilera:2016}. On the other hand, if the planet possesses a rocky core and is fully ablated before this point, this could 
imply the existence a population of close-in, super-Earth remnant cores orbiting subgiant stars, a prediction that could be testable with the upcoming Transiting Exoplanet Survey Satellite (TESS) mission.  

More detailed theoretical studies are needed to provide a clearer picture of the future evolution of the KELT-9 system and its analogs, and provide testable predictions. Further follow-up observations using ground and space-based telescopes will test models of heat redistribution, radius inflation, unusual photochemistry, and rapid ablation of planetary atmospheres. 
The leap from WASP-33b to KELT-9b should invigorate further exploration of the planet population of even higher-mass host stars, complementing efforts to discover planets orbiting ever lower-mass host stars\cite{Gillon:2017}.
The KELT-9 system provides an important benchmark system for understanding the nature of planetary systems around massive stars, from birth to death.


\begin{addendum} 
 \item[Acknowledgments] This research was made possible by the KELT survey, the KELT Follow-Up Network, and support from The Ohio State University, Vanderbilt University, and Lehigh University.
Work by B.S.G. and D.J.S was partially supported by NSF
CAREER Grant AST-1056524. 
K.G.S. and K.A.C. acknowledge partial support from NSF PAARE grant AST-1358862.
B.S.G. acknowledges support by the Jet Propulsion Laboratory, operated by the California Institute of Technology, and the Exoplanet Exploration Program of the National Aeronautics and Space Administration (NASA). B.J.F. notes that this material
is based upon work supported by the National Science
Foundation Graduate Research Fellowship under grant No.
2014184874. Any opinion, findings, and conclusions or recommendations
expressed in this material are those of the authors(s)
and do not necessarily reflect the views of the National
Science Foundation.  Work performed by J.E.R. was supported by the
Harvard Future Faculty Leaders Postdoctoral fellowship. K.K.M. acknowledges
the purchase of SDSS filters for Whitin Observatory by the
Theodore Dunham, Jr., Grant of the Fund for Astronomical
Research. N.N. acknowledges support by Japan Society for Promotion of Science (JSPS) KAKENHI Grant Number JP25247026. We acknowledge observations by Masanobu Kunitomo, Ryo Hasegawa, Bun\'ei Sato, Hiroki Harakawa, Teruyuki Hirano, and Hideyuki Izumiura on the Okayama 188cm telescope HIDES observations and Nobuhiko Kusakabe, Masahiro Onitsuka, and Tsuguru Ryu for MuSCAT observations. 
The NIRC2 AO data in this work were obtained
at the W.M.Keck Observatory, which was financed by the
W.M. Keck Foundation and is operated as a scientific partnership
between the California Institute of Technology, the
University of California, and NASA. The authors wish to recognize and acknowledge the very significant cultural role and reverence that the summit of Mauna Kea has always had within the indigenous Hawaiian community.  We are most fortunate to have the opportunity to conduct observations from this mountain.
This work has made use of NASA’s Astrophysics Data
System, the Exoplanets Data Explorer at exoplanets.org, the Extrasolar Planet Encyclopedia at exoplanet.eu, the SIMBAD database operated at
CDS, Strasbourg, France, and the VizieR catalogue access
tool, CDS, Strasbourg, France.
This publication makes use of data products from the
Widefield Infrared Survey Explorer, which is a joint project
of the University of California, Los Angeles; the Jet
Propulsion Laboratory/California Institute of Technology,
which is funded by NASA; the Two Micron All Sky Survey, which is a
joint project of the University of Massachusetts and the Infrared
Processing and Analysis Center/California Institute of
Technology, funded by NASA; and the
American Association of Variable Star Observers (AAVSO)
Photometric All-Sky Survey (APASS), whose funding is
provided by the Robert Martin Ayers Sciences Fund and
the AAVSO Endowment (https://www.aavso.org/aavso-photometric-all-sky-survey-data-release-1).
We would like to acknowledge useful input from Travis Barman, Jonathan Fortney, Mark Marley, and Kevin Zanhle. We would particularly like to thank the referees, who provided comments that lead to a much improved paper.
 \item[Author Contributions] 
 B.S.G. led the process from initial candidate selection to final planet confirmation. B.S.G. and K.G.S. wrote the majority of the main manuscript and contributed to the analysis of the results.  
 K.A.C. principally coordinated the assemblage of the final results and production of the Methods. T.G.B. and G.Z. lead the Doppler Tomographic analysis, and also contributed to the final modeling and interpretation. 
 J.D.E., R.J.S., D.J.S., L.A.B., J.P., J.E.R., K.D.C., M.C.J., and M.P. all provided critical insight, analysis, or interpretation of the system.  G.D., V, B., S.C.N., M.T.D., T.E., C.G., H.J.-C., D.H.K, A.F., J.G., A.I., J.F.K., M.M., K.M., N.N., T.E.O., P.A.R., G.S., D.C.S., R.R.Y., R.Z., B.J.F., and A.H., all provided photometric or radial velocity data that was important for the interpretation of the system. J.C. and E.J.G. provided the observations and analysis of the Keck AO data.  E.L.N.J., D.J.J., D.B., I.A.C., D.L.D., G.A.E., A.G., M.D.J., R.B.K, J.L.-B., M.B.L, J.M. K.K.M., R.W.P, H.R. D.C.S., C.S., T.G.T, M.T, and P.T. have all been essential for the initiation and successful operation of the KELT-North and KELT-South surveys.  All of the authors have read both the main manuscript and supplemental material and concur with the conclusions therein.  
 \item[Author Information] The authors declare that they have no
competing financial interests.  Correspondence and requests for materials should be addressed to B.S.G.~(email: gaudi.1@osu.edu).
\end{addendum}

\clearpage

\newcommand\mynotes[1]{\textcolor{red}{#1}}
\newcommand{\tick}{\ding{52}}
\newcommand{\cross}{\ding{53}}
\newcommand{\MS}{M\textsubscript{$\odot$}}    
\newcommand{\RS}{R\textsubscript{$\odot$}}    
\newcommand{\LS}{L\textsubscript{$\odot$}}    
\newcommand{\MJ}{M\textsubscript{J}}          
\newcommand{\RJ}{R\textsubscript{J}}          
\newcommand{\MP}{M\textsubscript{P}}          
\newcommand{\RP}{R\textsubscript{P}}          
\newcommand{\ME}{M\textsubscript{$\oplus$}}   
\newcommand{\RE}{R\textsubscript{$\oplus$}}   
\newcommand{\Mstar}{M\textsubscript{$\star$}} 
\newcommand{\Rstar}{R\textsubscript{$\star$}} 
\newcommand{\newln}{\\&\quad\quad{}}
\newcommand{\kms}{\,km\,s$^{-1}$}
\newcommand{\pchisq}{\ensuremath{P\left(>\chisq \right)}}
\newcommand{\chisq}{\ensuremath{\chi^{\,2}}}
\newcommand{\chisqr}{\ensuremath{\chi^{\,2}_{r}}}
\newcommand{\multifast}{{\tt MULTIFAST}}

\newcommand{\nodata}{...}

\methods
\setcounter{figure}{0}    
\setcounter{table}{0}
\renewcommand{\tablename}{Extended Data Table}


\section*{KELT-North Observations and Photometry\label{sec:keltobs}}

KELT-North field 11 is centered on $\alpha =$ 19$^{h}$ 27$^{m}$ 00$^{s}$, $\delta =$ 31$\degr$ 39$\arcmin$ 56$\farcs$16 (J2000) and was observed 6001 times from UT 2007 May 30 to UT 2013 June 14. Following the standard KELT candidate selection strategy\cite{Siverd:2012}, we reduced the data and extracted $\sim 116,000$ light curves from the east orientation and $\sim 138,000$ light curves from the west. The combined east and west light curves that passed the reduced proper motion cut\cite{Collier:2007} were searched for transiting exoplanet candidates. One bright ($V\sim7.6$) candidate, KC11C043952 (HD 195689, TYC 3157-638-1, 2MASS J20312634+3956196) located at $\alpha =$ 20$^{h}$ 31$^{m}$ 26$\fs$35401, $\delta =$ 39$\degr$ 56$\arcmin$ 19$\farcs$7744 (J2000), robustly passed our 
selection criteria\cite{Siverd:2012} 
making it a top candidate. Hereafter, we refer to the candidate host star as ``KELT-9'' and the candidate planet as ``KELT-9b''. KELT-9 is also located in KELT-North field 12, which is centered on $\alpha =$ 21$^{h}$ 22$^{m}$ 52$\fs$8, $\delta =$ 31$\degr$ 39$\arcmin$ 56$\farcs$16 (J2000). The field was observed 5,700 times from UT 2007 June 08 until UT 2013 June 14. 
Although not originally used to select KELT-9 as a candidate, the signal in the light curve from field 12 bolstered KELT-9 as a strong candidate.

The KELT-North field 11 phased KELT-9 discovery light curve is shown in the top panel of Fig. 1 in the main manuscript.
The field 11 light curve shows apparent out-of-transit variations (OOTVs), while these are absent in the field 12 light curve. We find that the source of the field 11 OOTVs is due to variability, possibly caused by saturation, of a star a few arc-minutes southeast of KELT-9 that is blended in the field 11 KELT-9 aperture. Given the magnitude of the variability, we suspect the contaminating source is the bright object HD 195728, rather than one of the fainter neighbors. The variation is long-term and not periodic, so it tends to inject power at and near aliases of our sampling rate. The variability from the neighbor does not affect the field 12 light curve because the point spread function of KELT-9 in field 12 is smaller and is elongated in a different direction than in field 11.  As a result, KELT-9 is minimally blended with the bright neighbor HD 195728, as well as the other faint neighbors near HD 195728.  

In short, the apparent out-of-transit variations (OOTVs) seen in the field 11 light curve are not due to intrinsic variability of KELT-9 itself, and thus do not lead us to question its reality as a bona fide planet-host candidate. 

Following the discovery of the primary transit signal, we carried out an intensive photometric followup campaign through which we obtained a total of 17 primary and 7 secondary transit light curves. These will be described in a forthcoming paper (Collins et al.\ 2017, in preparation), and are used in the global fit of the system presented below.  The combined, binned follow-up light curve is shown in Figure 1 of the main manuscript in order to highlight the statistical power of this follow-up dataset.  We only fit to the individual data; the binned data are shown only for illustration.  

\section*{Spectroscopic Follow-up\label{sec:spectroscopy}} 

To constrain the mass and enable eventual Doppler tomographic (DT) detection of KELT-9b, we obtained a total of 115 spectroscopic observations of the host star with the Tillinghast Reflector Echelle Spectrograph (TRES) on the 1.5\,m telescope at the Fred Lawrence Whipple Observatory, Arizona, USA. Each spectrum delivered by TRES has a spectroscopic resolution of $\lambda / \Delta \lambda = 41000$ over the wavelength range of $3900-9100$\,\AA\ over 51 echelle orders. This includes 40 observations covering the entire orbital phase to constrain the mass of the planet, and 75 observations made in-transit over three epochs to perform the tomographic line profile analysis. We measured the relative radial velocity from 104 of the observations (see Extended Data Table \ref{tab:rv1}) and used a total of 43 out-of-transit RVs (40 plus one out-of-transit RV from each of the spectroscopic transit observations) to constrain the planet's orbit and mass. The phased radial velocities are displayed in Figure 1 of the main article. 

Three spectroscopic transits of KELT-9 were observed with TRES on 2014-11-15, 2015-11-06, and 2016-06-12. The line broadening kernel is derived from each spectrum via a least-squares deconvolution\cite{Donati:1997,Collier:2010a,Zhou:2016}. The planetary shadow is seen crossing the stellar surface on all three nights, as shown in Extended Data Figure \ref{fig:dt3}. In addition, the rotational profiles allowed us to accurately determine a rotational velocity of $\vsini = 111.4\pm 1.3\,\mathrm{km\,s}^{-1}$. 

\section*{Host Star Properties\label{sec:star_props}}

Extended Data Table \ref{tbl:Host_Lit_Props} lists various properties and measurements of KELT-9 collected from the literature and derived in this work. The data from the literature include the four monochromatic near-UV fluxes from the Catalog of Stellar UV Fluxes\cite{Thompson:1995}, $UBV$ photometry\cite{Mermilliod:1991},
optical fluxes in the $B_{\rm T}$ and $V_{\rm T}$ passbands from the Tycho-2 catalog\cite{Hog:2000}, 
$I_{\rm C}$ from the TASS catalog\cite{Droege:2006},
near-infrared (IR) fluxes in the $J$, $H$ and $K_{\rm S}$ passbands from the 2MASS Point Source Catalog\cite{Cutri:2003,Skrutskie:2006}, near- and mid-IR fluxes in four WISE passbands\cite{Wright:2010,Cutri:2012}, distances from Hipparcos\cite{Leeuwen:2007} and Gaia\cite{Gaia:2016}, and proper motions from the NOMAD catalog\cite{Zacharias:2013,Zacharias:2004}.


\section*{SED Analysis}

\begin{figure}
\centering
\includegraphics[width=\linewidth,trim=3.0cm 2.2cm 3.0cm 3.2cm, clip=true]{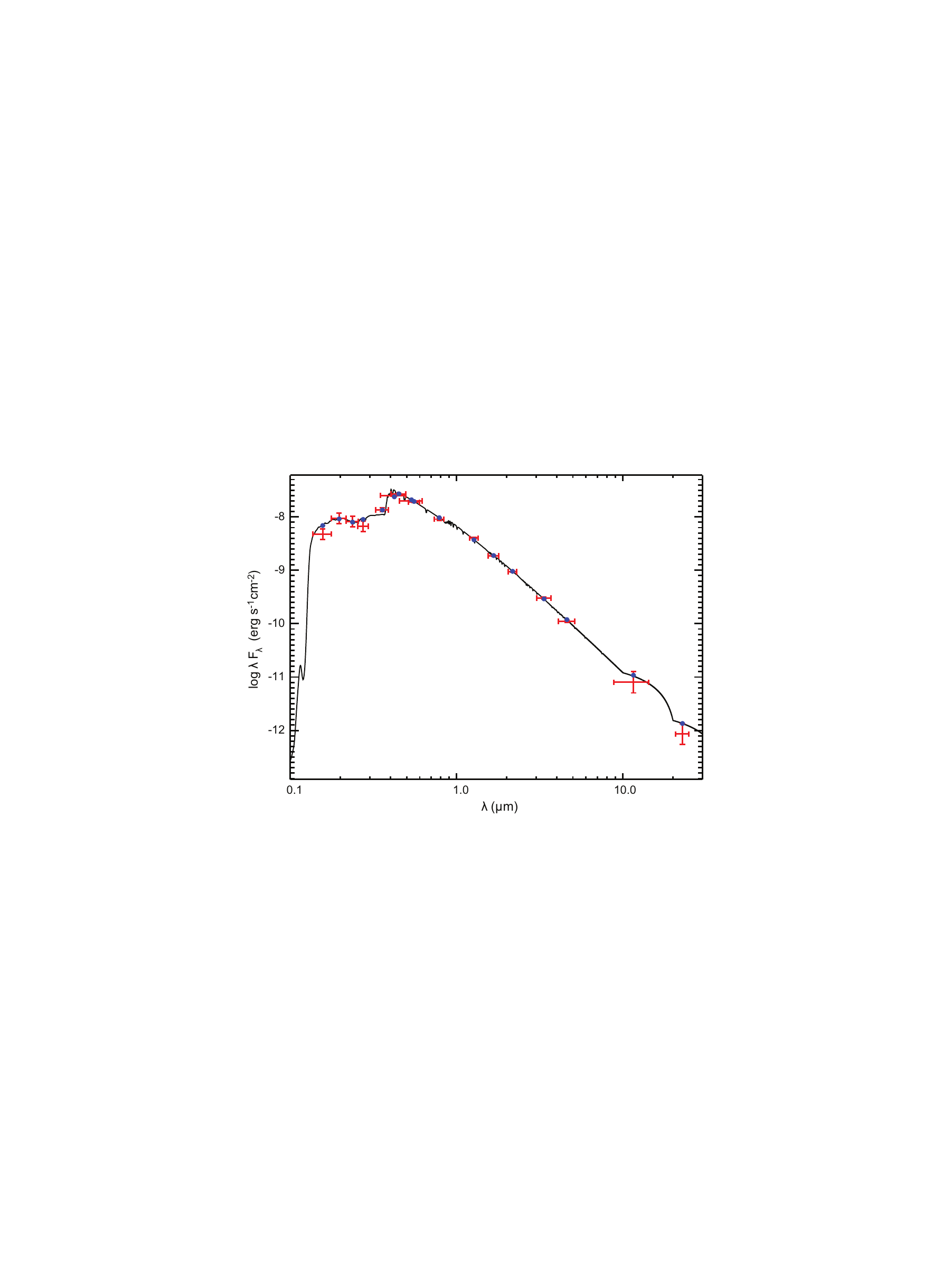}
\vskip-3.5in
\caption{KELT-9 spectral energy distribution (SED). Crosses represent the measured fluxes, with vertical bars representing the measurement uncertainties and the horizontal bars representing the width of the bandpass. The blue dots are the predicted passband-integrated fluxes of the best-fit theoretical SED corresponding to our observed photometric bands. The black curve represents the best-fit Kurucz stellar atmosphere\cite{Kurucz:1992}.}
\label{fig:sed}
\end{figure}

We construct an empirical, broad-band spectral energy distribution (SED) of KELT-9,
shown in Extended Data Figure \ref{fig:sed}.  We use the 17 photometric measurements from the literature discussed in Section \ref{sec:star_props} and shown in Extended Data Table \ref{tbl:Host_Lit_Props}. In total, the observed SED spans the wavelength range 0.16--22~$\mu$m. We fit this observed SED to Kurucz stellar atmosphere models\cite{Kurucz:1992}. 
For simplicity we adopted a fixed $\loggstar = 4.0$, based on the light curve transit analysis. The fit parameters were thus the effective temperature ($\teff$), the metallicity ($\feh$), the extinction ($A_V$), and the overall flux normalization. The maximum permitted extinction was set to $A_V = 0.30$ based on the total line-of-sight extinction in the direction of KELT-9 from Galactic dust maps\cite{Schlegel:1998}. 

\section*{Stellar Parameters from SED\label{sec:star_pars_sed}}

The best fit model has a reduced $\chi^{2}$
of 2.56 for 13 degrees of freedom (17 flux measurements, 4 fit parameters).  We find $A_{\rm V}$ = $0.09 \pm 0.05$, $\feh = 0.0 \pm 0.2$, and $\teff = 9560 \pm 550$~K. 
We note that the quoted statistical uncertainties on $A_{\rm V}$ and $\teff$ are
likely to be slightly underestimated because we have not accounted for the
uncertainty in the value of $\loggstar$ used to derive the model SED, although this parameter generally does not strongly affect the overall shape of the SED, and moreover \loggstar\ is strongly constrained from the light curve transit analysis. 

We can integrate the best-fit SED to obtain the (unextincted) bolometric flux at Earth, $\fbol = 3.18(\pm 0.09) \times 10^{-8}$ erg~s$^{-1}$~cm$^{-2}$. Together with the best-fit $\teff$ and the distance newly provided by the {\it Gaia} parallax, we obtain a direct constraint on the stellar radius of $\Rstar = 2.37 \pm 0.35$~\rsun. From the distance provided by the Hipparcos parallax, we obtain a direct constraint\cite{Stassun:2016} on the stellar radius of $\Rstar = 2.17 \pm 0.33$~\rsun.   
As noted in the footnote of Extended Data Table~\ref{tbl:Host_Lit_Props}, 
we generally believe the Hipparcos-derived distance and stellar radius to currently be more reliable, and use this for most of the analysis in this paper.

\section*{Stellar Models and Age\label{sec:hrd_and_age}}

\begin{figure}
\centering
\includegraphics[width=\linewidth,trim=3.0cm 2.2cm 3.0cm 3.2cm, clip=true]{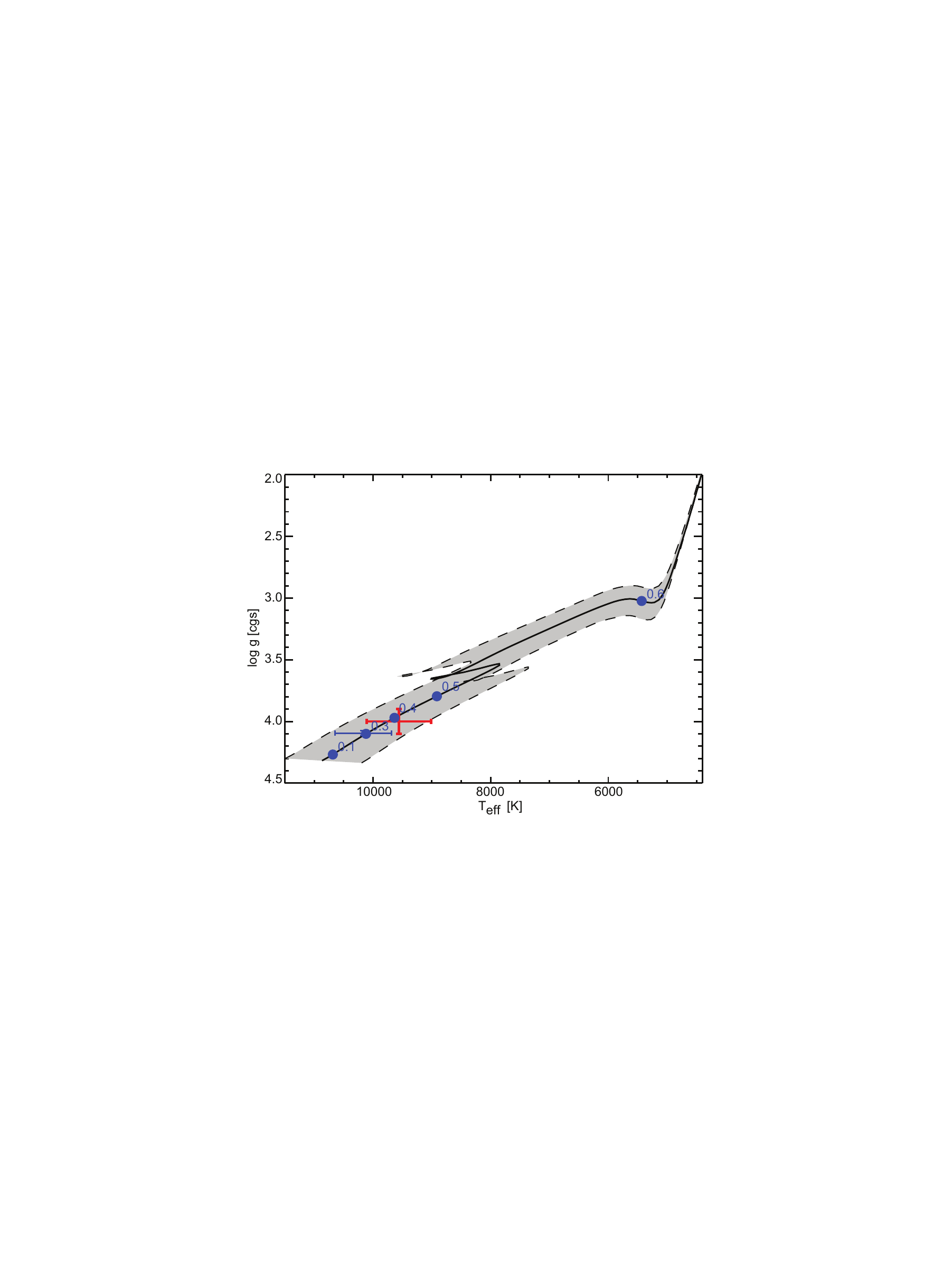}
\vskip-3.5in
\caption{Evolution of the KELT-9 system in the Hertzsprung-Russell diagram. The red cross represents the system parameters from the initial SED fit, the blue cross represents the final system parameters from the global fit. The black curve represents the theoretical evolutionary track for a star with the mass and metallicity of KELT-9, and the grey swath represents the uncertainty on that track based on the uncertainties in mass and metallicity. Nominal ages in Gyr are shown as blue dots.
When KELT-9 evolves to the base of the giant branch in 200-300 million years, it will encroach upon the orbit of KELT-9b.  The fate of the system at that point is highly uncertain \cite{Anguilera:2016,Metzger:2012,Kunitomo:2011,Privitera:2016,Villaver:2007}.
}
\label{fig:hrd}
\end{figure}

With $\teff$ and $\loggstar$, and an estimated stellar mass from the global analysis (see below), we can place the KELT-9 system in the Hertzsprung-Russell diagram for comparison with theoretical stellar evolutionary models (Extended Data Figure~\ref{fig:hrd}). 
From the values obtained with the initial SED fit, we infer a system age of $\approx$0.4~Gyr; the final age estimate using the final global fit parameters is $\approx$0.3~Gyr. Thus, it is clear that the KELT-9 system is nearly unevolved from the zero-age main-sequence, and in any event is at an early stage of evolution well before the ``blue hook" transition to the subgiant and eventual red giant evolutionary phases. 

\section*{Global System Fit\label{sec:globalfit}}

We determined the physical and orbital parameters of the KELT-9 system by jointly fitting 17 primary and 7 secondary light curves, 43 TRES out-of-transit RVs, and 3 Doppler tomographic data sets (see the section Doppler tomographic model below). To perform the global fit, we used {\tt MULTI-EXOFAST} (\multifast\ hereafter), which is a custom version of the public software package {\tt EXOFAST}\cite{Eastman:2013}. \multifast\ first performs an {\tt AMOEBA}\cite{Nelder:1965} best fit to each of the RV and light curve data sets individually to determine uncertainty scaling factors. The uncertainties are scaled such that the probability that the \chisq \ for a data set is larger than the value we achieved, $\pchisq$, is $0.5$, to ensure the resulting parameter uncertainties are roughly accurate. 
The resulting RV uncertainty scaling factor is 1.23. 
The DT uncertainties were scaled based on the \chisq\ of the out-of-transit data relative to the median value. The uncertainties of the UT 2014-11-15, UT 2015-11-06, and UT 2016-06-12 DT observations were scaled by 0.79, 0.78, and 0.79, respectively. Finally, \multifast\ performs a joint {\tt AMOEBA} model fit to all of the datasets and executes Markov Chain Monte Carlo (MCMC), starting at the global best fit values, to determine the median and 68\% confidence intervals for each of the physical and orbital parameters. Siverd et al.\cite{Siverd:2012} provide a more detailed description of \multifast, except the Doppler tomographic model implementation, which was newly implemented as part of this work.

\section*{Doppler Tomographic Model\label{sec:dopptom}}

To model the Doppler tomographic signal, we construct and integrate our own models into the \multifast\ fitting process\cite{Boue:2013}. We treat the Doppler shadow of the planet as a combination of three different broadening mechanisms which we account for consecutively. Our shadow model begins as a Gaussian profile, with a standard deviation equal to the mean inherent spectral line width (in velocity space) of an equivalent non-rotating star. Second, we convolve this base Gaussian profile with a second Gaussian of width $\sigma=c/R$ to account for the finite spectral resolution of the TRES spectrograph ($R\approx44,000$). Third, and finally, we convolve the resulting shadow model with a normalized rotational broadening kernel. 

For our rotational broadening kernel we use the kernel given by Equation 18.14 in Gray\cite{Gray:2008}. For simplicity we assume that the linear limb-darkening coefficient ($\epsilon$) for the kernel is equal to zero. We set the velocity width of the kernel equal to $(R_p/R_*)\,\vsini$, to represent the fraction of the stellar rotational surface actually obscured by the planet. 

\begin{figure}
\includegraphics[width=1.0\linewidth]{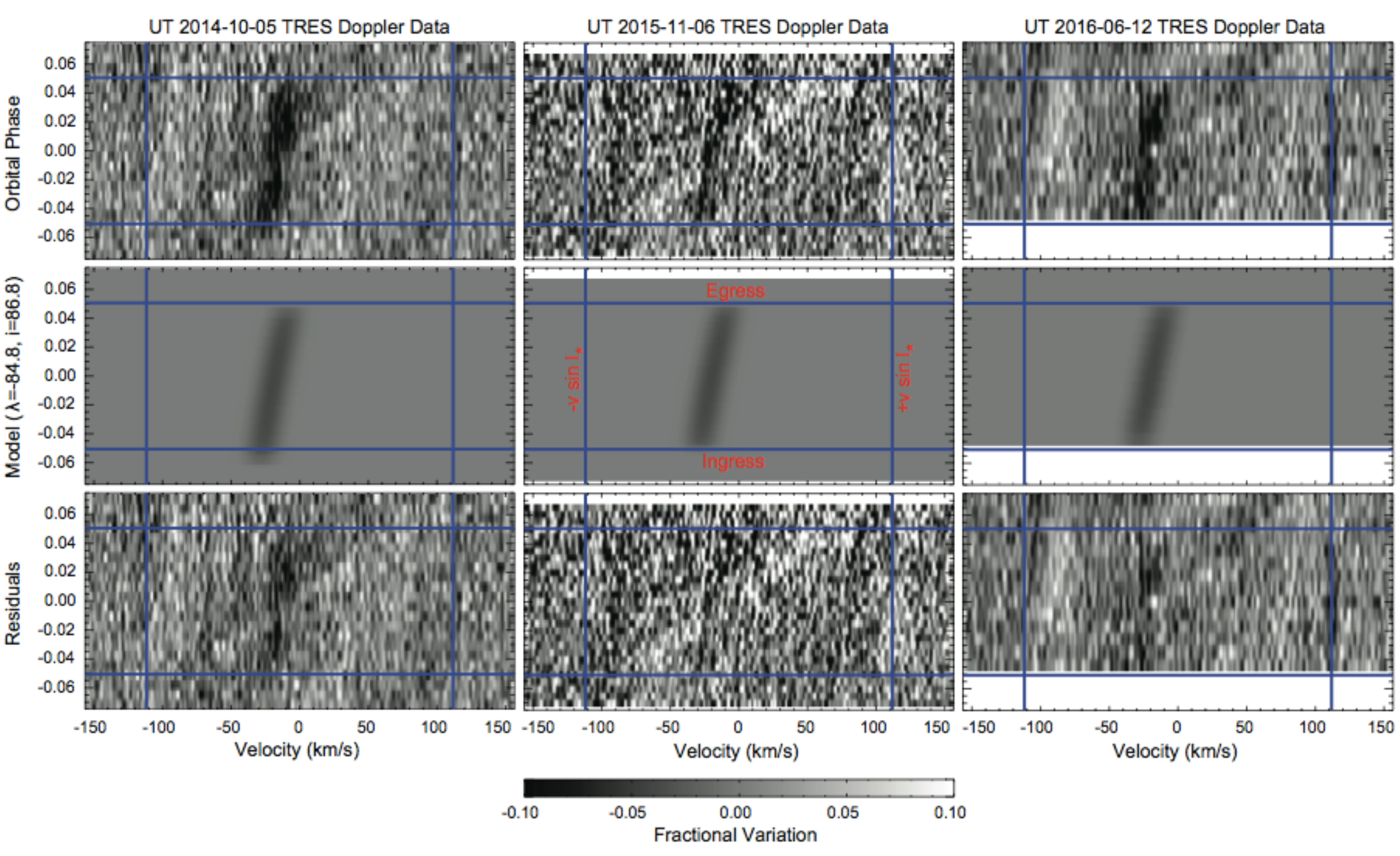}
\caption{Doppler tomographic line profile plots.  The data, models, and residuals from all three nights where KELT-9 was observed during transit are shown in three columns.  In each case, the top plot shows the spectroscopic date, the middle panels show the derived model, and the bottom panels show the residuals.  In each panel, the vertical blue lines denote the width of the convolution kernel (i.e., $\vsini$), and the horizontal blue lines show the duration of the transit. Time increases vertically for each panel. The apparent extension of the Doppler shadow before ingress on UT 20014-10-05 is an artifact of uneven and widely-spaced sampling in time. The greyscale shows the fractional variation in the spectroscopic signal from the null hypothesis of no shadow due to a transiting planet. Darker regions indicate the Doppler shadow as the planet crosses the face of the host star.  Note the transit is nearly coincident with the projected stellar spin axis, implying a nearly polar orbit for the planet. 
\label{fig:dt3}}
\end{figure}

We normalize the resulting shadow model using the depth of a $V$-band transit light curve for the system at each observation time, so that the integrated light obscured by the shadow matches the actual depth of the transit. To calculate the trajectory of the shadow in time-velocity space, we follow the method given by Equations 7, 8, and 10 in Collier Cameron et al.\cite{Collier:2010b}  

Besides the physical parameters already included in the transit model, our Doppler tomography model therefore has three free parameters: the rotation velocity of the host star ($\vsini$), the inherent line width of the non-rotating stellar spectrum ($\sigma_0$), and the spin-orbit angle ($\lambda$). Since KELT-9 has $\vsini = 111.4\pm 1.3\,\mathrm{km\,s}^{-1}$, $\sigma_0$ is essentially negligible, so we use a prior on $\sigma_0 = 10.0\pm 1.0\,\mathrm{m\,s}^{-1}$ (i.e. a value near zero that avoids significant numbers of negative $\sigma_0$ MCMC trials) to improve convergence of the MCMC chains. To evaluate the goodness-of-fit for a particular set of model parameters, we use the $\Delta\chi^2$ between the predicted shadow model and the observed Doppler tomographic observations. Since the observations themselves are super-sampled below the instrumental resolution of the spectrograph during the data reduction process, we divide the resulting $\Delta\chi^2$ value by a factor of $(c/R)/\Delta v$, where $c$ is the speed of light, $R$ is the instrumental spectral resolution, and $\Delta v$ is the velocity interval between the individual points in the super-sampled spectra. This reduction of the $\Delta\chi^2$ accounts for the fact that due to the instrumental spectral resolution, adjacent points in the super-sampled spectra were not truly independent.

\section*{Global Model Results\label{sec:globalresults}}

We adopt a fiducial model (Model 1) with YY constraints, a fixed circular orbit, a fixed RV slope $\dot{\gamma} = 0$, and a Hipparcos-based \Rstar\ prior, and compare the results to those from ten other global models that systematically explore the results of differing constraints. The posterior median parameter values and 68\% confidence intervals are shown in Extended Data Table \ref{tab:parameters} for the initial TTV and final ten global model fits. The KELT-9 fiducial model indicates the system has a host star with mass $\Mstar = 2.52 \msun$, radius $\Rstar = 2.362 \rsun$, and effective temperature $\teff = 10,170$~K, and an extremely hot planet with $\teq=4050$~K, mass $\MP = 2.88 \MJ$, and radius $\RP = 1.891 \mj$.

The initial TTV model (Model 0) parameter median values are well within $1\sigma$ of the fiducial model, and the uncertainties are essentially identical to those of the fiducial model results, except the ephemeris parameter uncertainties ($T_C$ and $P$) are $\sim$ 25\% larger due to the additional seven secondary transits in the fiducial model and the linear ephemeris constraint.

We also explored several other models involving different choices of observational constraints and free-fit parameters, which will be described in a forthcoming paper (Collins et al.\ 2017, in preparation) but are summarized in Extended Data Table~\ref{tab:parameters}.

In summary, we find that all combinations of stellar constraints result in system parameter values that are within $1\sigma$, and in almost all cases, well within $1\sigma$ of the fiducial model.

\section*{False Positive Analysis\label{sec:false}}

KELT-9 was a sufficiently unusual system that we were especially vigilant about ruling out false positive scenarios.
We have many lines of evidence that rule out essentially all viable false positive scenarios.  First, false positives around rapidly rotating stars can be ruled out via relatively imprecise RV measurements, high precision light curves, and a positive Rossiter-McLaughlin\cite{Rossiter:1924,McLaughlin:1924} (RM) or DT detection\cite{Collier:2010a,Bieryla:2015}.  An upper limit on the Doppler signal with relatively imprecise RV precisions of a few hundred ${\rm m~s^{-1}}$ can rule out low-mass stars and brown dwarfs as the occulter of the primary star.  Precise follow-up light curves, along with the \vsini\ measured from the spectra, can be used to predict the magnitude, impact parameter, and duration of the DT signal (although not the direction $\lambda$). Thus a measurement of the DT signal that is consistent with the light curve effectively confirms the planet interpretation.  

In the case of KELT-9, our first spectroscopic measurement during transit yielded a very weak RM signal, which was sufficiently noisy that we considered it could simply be due to the relatively large uncertainties of the RVs.  This lack of a RM signal would be surprising for a transiting planet given the transit depth and large \vsini\ of the star.  We thus originally concluded that the system was likely a false positive.  However, we decided to proceed with a full DT analysis, which showed tentative evidence for a planet shadow that was nearly coincident with the projected stellar rotation axis, thus possibly explaining the small RM signal.  Improved reduction methods and two additional DT measurements definitively showed that the planet does indeed transit nearly along the projected stellar rotation axis, as shown in Extended Data Figure \ref{fig:dt3}. 

Second, we were also concerned about the lack of a definitive measurement of the Doppler reflex signal.  Our initial fits to a subset of the data presented here provided only upper limits to $K$, and thus we were originally not able to measure the planet mass.  However, after acquiring additional data, we were ultimately able to measure $K$ to roughly $\sim 30\%$ precision.  Indeed, a Lomb-Scargle\cite{Lomb:1976,Scargle:1982} periodogram of the out-of-transit RV data yielded several significant signals.  The two signals with the highest power are at long periods of $\sim 100$ and $\sim 200$ days (depending on the exact choice of datasets that are used), which are likely aliases of each other, and may be due to another planet, intrinsic long-term stellar activity, or may simply be due to systematic errors in the RVs.  However, the signal with the next highest power has a ephemeris of $T_{\rm 0}=2456969.7901$ and $P=1.4811228$.  This period is consistent with the final period we derive from our follow-up light curves to within $\sim 10^{-5}$~d, and projecting the ephemeris from our follow-up light curves forward yields a value of $T_{\rm 0}$ that is consistent with that from the RV periodogram to within $\sim 10^{-4}$~d. We are therefore confident that the reflex RV signal is real and due to the transiting planet. 

Finally, there are several other pieces of evidence that support the planet hypothesis.  (1) The stellar density we infer from the global fit to the light curve (which is essentially a direct observable) is consistent with what one would expect for an unevolved A0 star with the spectroscopically-measured temperature. (2) The limb darkening in the redder bandpasses are noticeably smaller by eye than planet transits of a cooler star, as one would expect for a hot star and bandpasses near the Rayleigh-Jeans part of the SED.  Although we do not fit for the limb darkening, the fact that the light curves spanning from $U$ to $z$ are well-fit by the smaller limb darkening coefficients that are predicted for a star of this temperature and surface gravity (as compared to, e.g., those expected for a solar-type star) provides quantitative support for this qualitative conclusion. (3) Adaptive optics observations (Collins et al.\ 2017, in preparation) do not reveal a blended stellar companion that could cause a false positive, down to the mass of $\sim 0.14~\msun$ with a projected separation of $\gtrsim 200$ AU.  (4) We detect a secondary eclipse in the $z$ band with a depth that is consistent with what one would expect given the amount of stellar irradiation that the planet receives. (5) The basic consistency between the model fits using various constraints (YY, Torres, and Hipparcos and Gaia-inferred radii) also provides support for our interpretation.

We therefore conclude that, despite the very unusual and extreme nature of the system, all available data are consistent with the interpretation that KELT-9 is being transited by an extremely irradiated, highly inflated planet on a near-polar orbit of only $\sim 1.5$~d.

\begin{addendum}
\item[Data Availability] The data that support the findings of this study are available from the corresponding
author upon reasonable request. 
\end{addendum}

\begin{figure}
\includegraphics[width=\linewidth]{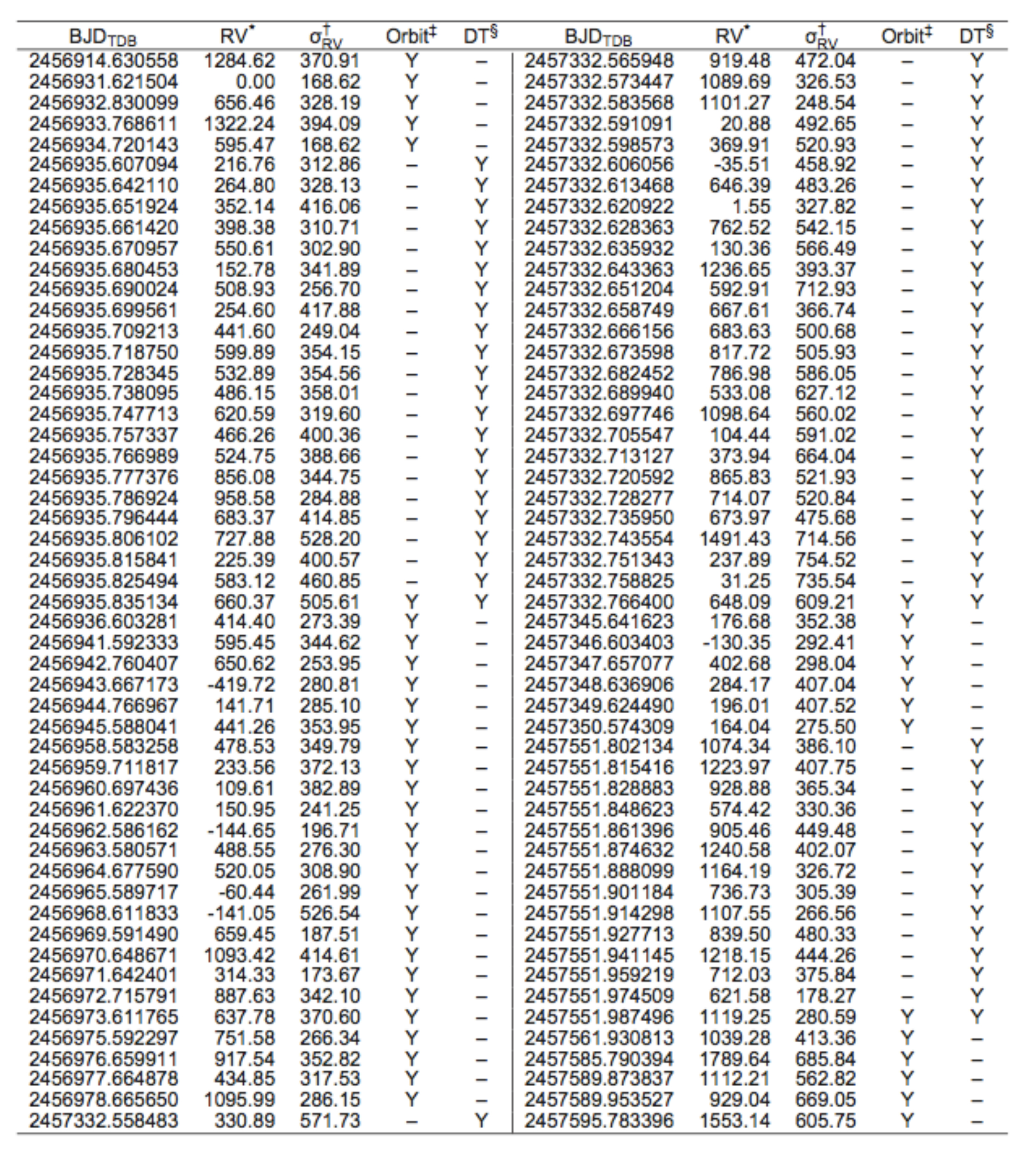}
\label{tab:rv1}
{\bf Extended Data Table 1:} Radial Velocity Measurements of KELT-9\newline
$^*$relative RVs (m s$^{-1}$)\newline
$^\dagger$unrescaled relative RV errors (m s$^{-1}$)\newline
$^\ddagger$Y indicates that the RV was included in the RV orbit analysis\newline
$^\S$Y indicates that the spectrum was included in the Doppler tomography analysis
\end{figure}

\begin{figure}
\includegraphics[width=\linewidth]{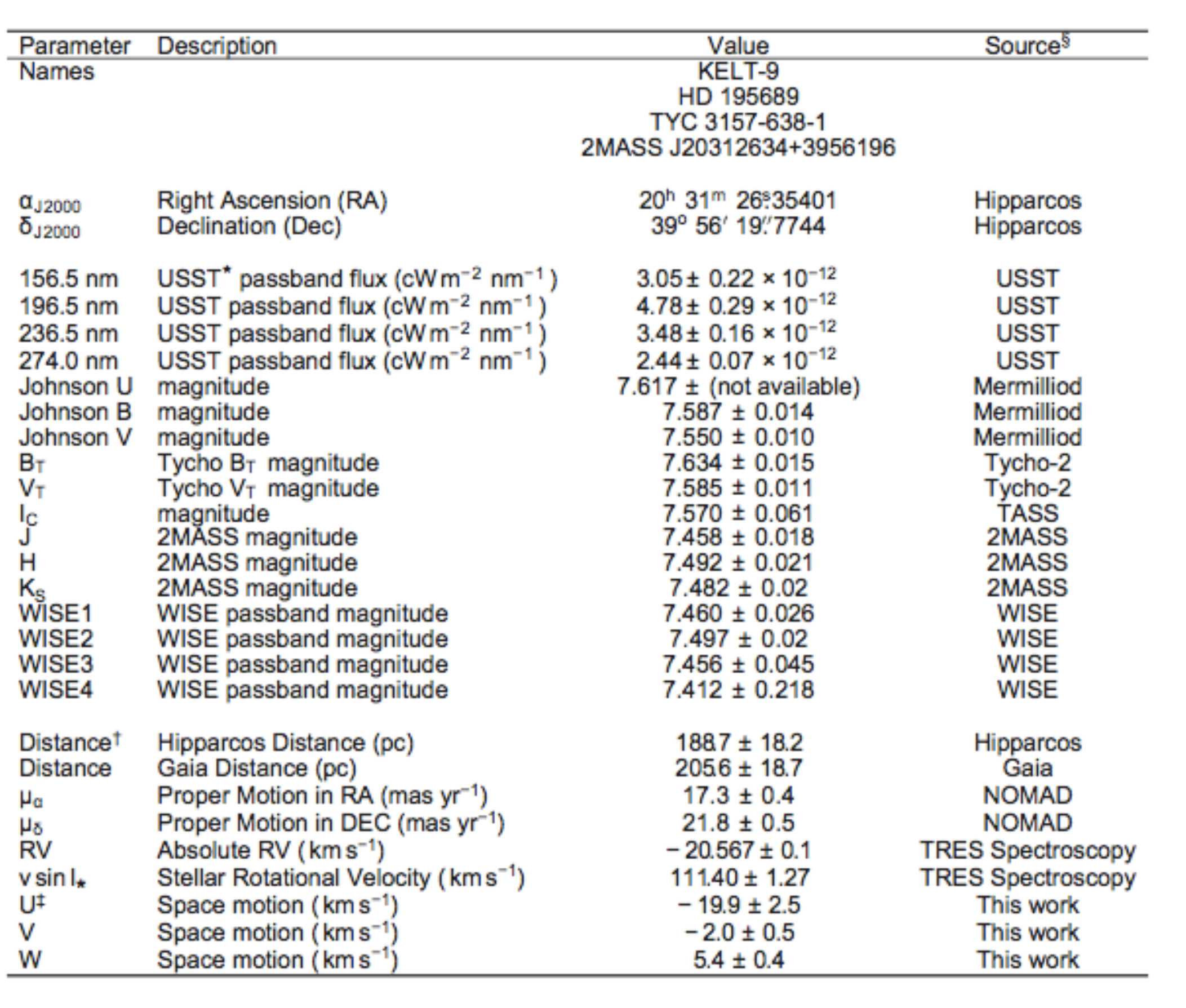}
\label{tbl:Host_Lit_Props}
{\bf Extended Data Table 2:} Stellar Properties of KELT-9 Obtained from the Literature and This Work\newline
$^*$Ultraviolet Sky Survey Telescope\newline
$^\dagger$We generally use the Hipparcos parallax as the default value for most the analysis in this paper, because the Hipparcos and Gaia parallaxes have similar fractional precisions, they agree to within their uncertainties, and the Gaia parallaxes are known to have a small systematic error, although this error is small compared to the absolute value of the parallax of KELT-9.  Nevertheless, when warranted, we compare results obtained using both parallax determinations.\newline
$^\ddagger$U is positive in the direction of the Galactic Center.\newline
$^\S$See text for references.
\end{figure}

\begin{figure}
\includegraphics[width=\linewidth]{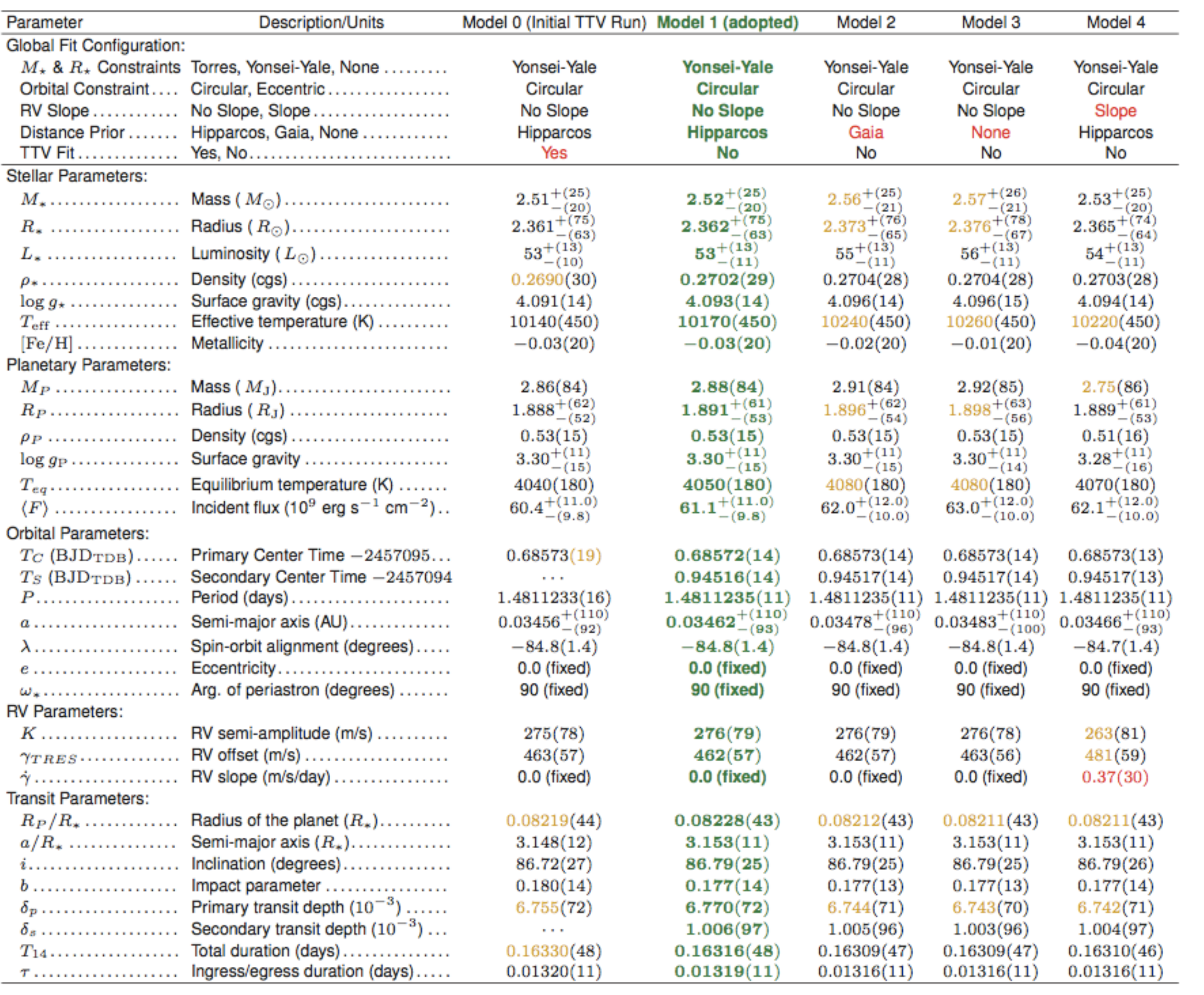}
\label{tab:parameters}
{\bf Extended Data Table 3:} Median Values and 68\% Confidence Intervals for Selected Physical and Orbital Parameters of the KELT-9 System\newline
The values within parentheses represent the uncertainty in the same number of least significant digits. For readability, we report asymmetric uncertainties with a difference of less than 5\% in the absolute value of the positive and negative component as a symmetric uncertainty with a value equal to the maximum of the absolute values of the asymmetric components. Green text highlights the adopted fiducial model values. Yellow text highlights nominal parameter values that differ slightly (but are well within $1\sigma$) from the fiducial model values. Red text highlights parameter nominal values and uncertainties that differ significantly (on the order of $1\sigma$) from the fiducial model values. Red text in the Global Fit Configuration section highlights configuration settings that differ from the fiducial model settings.
\end{figure}

\begin{figure}
\includegraphics[width=\linewidth]{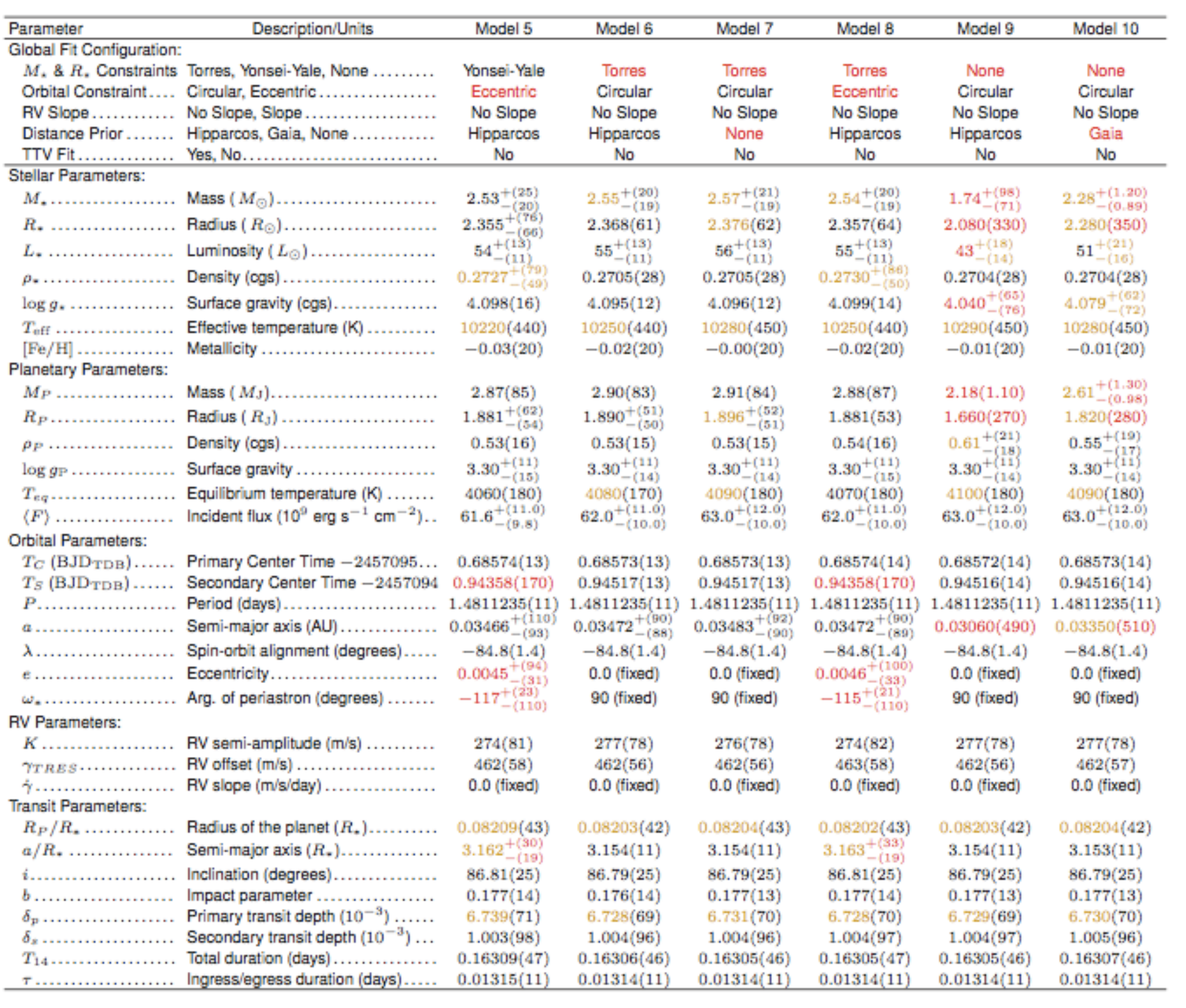}
{\bf Extended Data Table 3:} Continued.
\end{figure}

\noindent{\bf References}

\begin{thebibliography}{1}
\bibitem{Collier:2010a} {{Collier Cameron}, A., {\it et al}.} {Line-profile tomography of exoplanet transits - II. A gas-giant planet transiting a rapidly rotating A5 star}. {\it Monthly Notices of the Royal Astronomical Society} {\bf 407}, {507-514} (2010). 
\bibitem{Vonessen:2015} {{von Essen}, C., {Mallonn}, M., {Albrecht}, S., {Antoci}, V., {Smith}, A.~M.~S., {Dreizler}, S., {Strassmeier}, K.~G.} {A temperature inversion in WASP-33b? Large Binocular Telescope occultation data confirm significant thermal flux at short wavelengths}. {\it \aap} {\bf 584}, A75 (2015). 
\bibitem{Komacek:2016} {{Komacek}, T.~D., {Showman}, A.~P.} {Atmospheric Circulation of Hot Jupiters: Dayside-Nightside Temperature Differences}. {\it \apj} {\bf 821}, 16 (2016). 
\bibitem{Demory:2011} {{Demory}, B.-O., {Seager}, S.} {Lack of Inflated Radii for Kepler Giant Planet Candidates Receiving Modest Stellar Irradiation}. {\it Astrophysical Journals} {\bf 197}, 12 (2011). 
\bibitem{Pecaut:2013} {{Pecaut}, M.~J., {Mamajek}, E.~E.} {Intrinsic Colors, Temperatures, and Bolometric Corrections of Pre-main-sequence Stars}. {\it \apjs} {\bf 208}, 9 (2013). 
\bibitem{Murray-Clay:2009} {{Murray-Clay}, R.~A., {Chiang}, E.~I., {Murray}, N.} {Atmospheric Escape From Hot Jupiters}. {\it \apj} {\bf 693}, {23-42} (2009). 
\bibitem{Charbonneau:2000} {{Charbonneau}, D., {Brown}, T.~M., {Latham}, D.~W., {Mayor}, M.} {Detection of Planetary Transits Across a Sun-like Star}. {\it Astrophysical Journal Letters} {\bf 529}, {L45-L48} (2000). 
\bibitem{Henry:2000} {{Henry}, G.~W., {Marcy}, G.~W., {Butler}, R.~P., {Vogt}, S.~S.} {A Transiting ``51 Peg-like'' Planet}. {\it Astrophysical Journal Letters} {\bf 529}, {L41-L44} (2000). 
\bibitem{Hartman:2015} {{Hartman}, J.~D., {\it et al}.} {HAT-P-57b: A Short-period Giant Planet Transiting a Bright Rapidly Rotating A8V Star Confirmed Via Doppler Tomography}. {\it Astronomical Journal} {\bf 150}, 197 (2015). 
\bibitem{Zhou:2017} {{Zhou}, G., {\it et al}.} {HAT-P-67b: An Extremely Low Density Saturn Transiting an F-Subgiant Confirmed via Doppler Tomography}. {\it ArXiv e-prints} (2017). {1702.00106}.
\bibitem{Borucki:2010} {{Borucki}, W.~J., {\it et al}.} {Kepler Planet-Detection Mission: Introduction and First Results}. {\it Science} {\bf 327}, 977 (2010). 
\bibitem{Johnson:2007} {{Johnson}, J.~A., {\it et al}.} {Retired A Stars and Their Companions: Exoplanets Orbiting Three Intermediate-Mass Subgiants}. {\it \apj} {\bf 665}, {785-793} (2007). 
\bibitem{Lloyd:2011} {{Lloyd}, J.~P.} {''Retired'' Planet Hosts: Not So Massive, Maybe Just Portly After Lunch}. {\it \apjl} {\bf 739}, L49 (2011). 
\bibitem{Schlaufman:2013} {{Schlaufman}, K.~C., {Winn}, J.~N.} {Evidence for the Tidal Destruction of Hot Jupiters by Subgiant Stars}. {\it \apj} {\bf 772}, 143 (2013). 
\bibitem{Galland:2005} {{Galland}, F., {Lagrange}, A.-M., {Udry}, S., {Chelli}, A., {Pepe}, F., {Queloz}, D., {Beuzit}, J.-L., {Mayor}, M.} {Extrasolar planets and brown dwarfs around A-F type stars. I. Performances of radial velocity measurements, first analyses of variations}. {\it \aap} {\bf 443}, {337-345} (2005). 
\bibitem{Borgniet:2017} {{Borgniet}, S., {Lagrange}, A.-M., {Meunier}, N., {Galland}, F.} {Extrasolar planets and brown dwarfs around AF-type stars. IX. The HARPS southern sample}. {\it \aap} {\bf 599}, A57 (2017). 
\bibitem{Pepper:2007} {{Pepper}, J., {\it et al}.} {The Kilodegree Extremely Little Telescope (KELT): A Small Robotic Telescope for Large-Area Synoptic Surveys}. {\it Publications of the Astronomical Society of the Pacific} {\bf 119}, {923-935} (2007). 
\bibitem{Pepper:2012} {{Pepper}, J., {Kuhn}, R.~B., {Siverd}, R., {James}, D., {Stassun}, K.} {The KELT-South Telescope}. {\it Publications of the Astronomical Society of the Pacific} {\bf 124}, {230-241} (2012). 
\bibitem{Casewell:2015} {{Casewell}, S.~L., {\it et al}.} {Multiwaveband photometry of the irradiated brown dwarf WD0137-349B}. {\it \mnras} {\bf 447}, {3218-3226} (2015). 
\bibitem{Metzger:2012} {{Metzger}, B.~D., {Giannios}, D., {Spiegel}, D.~S.} {Optical and X-ray transients from planet-star mergers}. {\it \mnras} {\bf 425}, {2778-2798} (2012). 
\bibitem{Anguilera:2016} {{Aguilera-G{\'o}mez}, C., {Chanam{\'e}}, J., {Pinsonneault}, M.~H., {Carlberg}, J.~K.} {On Lithium-rich Red Giants. I. Engulfment of Substellar Companions}. {\it \apj} {\bf 829}, 127 (2016). 
\bibitem{Gillon:2017} {{Gillon}, M., {\it et al}.} {Seven temperate terrestrial planets around the nearby ultracool dwarf star TRAPPIST-1}. {\it Nature} {\bf 542}, {456-460} (2017). 
\bibitem{Eastman:2013} {{Eastman}, J., {Gaudi}, B.~S., {Agol}, E.} {EXOFAST: A Fast Exoplanetary Fitting Suite in IDL}. {\it Publications of the Astronomical Society of the Pacific} {\bf 125}, {83-112} (2013). 
\bibitem{Demarque:2004} {{Demarque}, P., {Woo}, J.-H., {Kim}, Y.-C., {Yi}, S.~K.} {Y$^{2}$ Isochrones with an Improved Core Overshoot Treatment}. {\it Astrophysical Journals} {\bf 155}, {667-674} (2004). 
\bibitem{Torres:2010} {{Torres}, G., {Andersen}, J., {Gim{\'e}nez}, A.} {Accurate masses and radii of normal stars: modern results and applications}. {\it Astronomy \& Astrophysics Rev.} {\bf 18}, {67-126} (2010). 
\bibitem{vanLeeuwen:2007} {{van Leeuwen}, F.} {Validation of the new Hipparcos reduction}. {\it \aap} {\bf 474}, {653-664} (2007). 
\bibitem{Gaia:2016} {{Gaia Collaboration}, {\it et al}.} {Gaia Data Release 1. Summary of the astrometric, photometric, and survey properties}. {\it Astronomy \& Astrophysics} {\bf 595}, {A2}, (2016)
\bibitem{Stassun:2016} {{Stassun}, K.~G., {Collins}, K.~A., {Gaudi}, B.~S.} {Accurate, Empirical Radii and Masses of Planets and their Host Stars with Gaia Parallaxes}. {\it Astronomical Journal} {\bf 153}, {136-155} (2017). 
\bibitem{Siverd:2012} {{Siverd}, R.~J., {\it et al}.} {KELT-1b: A Strongly Irradiated, Highly Inflated, Short Period, 27 Jupiter-mass Companion Transiting a Mid-F Star}. {\it Astrophysical Journal} {\bf 761}, 123 (2012). 
\bibitem{Johnson:2015} {{Johnson}, M.~C., {Cochran}, W.~D., {Collier Cameron}, A., {Bayliss}, D.} {Measurement of the Nodal Precession of WASP-33 b via Doppler Tomography}. {\it \apjl} {\bf 810}, L23 (2015). 
\end{thebibliography}

\noindent{\bf References}

\begin{thebibliography}{1}
\makeatletter
\addtocounter{\@listctr}{30}
\makeatother

\bibitem{Collier:2007} Collier Cameron, A. {\it et al.} Efficient identification of exoplanetary transit candidates from SuperWASP light curves. {\mnras}, {\bf 380}, 1230-1244 (2007). 

\bibitem{Donati:1997} {Donati}, J.-F., {Semel}, M., {Carter}, B.~D., {Rees}, D.~E., {Collier Cameron}, A. Spectropolarimetric observations of active stars. {\mnras}, {\bf 291}, 658 (1997). 

\bibitem{Zhou:2016} {Zhou}, G., {Latham}, D.~W., {Bieryla}, A., {Beatty}, T.~G., {Buchhave}, L.~A., {Esquerdo}, G.~A., {Berlind}, P., {Calkins}, M.~L. Spin-orbit alignment for KELT-7b and HAT-P-56b via Doppler tomography with TRES. {\mnras}, {\bf 460}, 3376-3383, (2016). 

\bibitem{Thompson:1995} {Thompson}, G.~I., {Nandy}, K., {Jamar}, C., {Monfils}, A., {Houziaux L.}, {Carnochan}, D.~J., {Wilson}, R. VizieR Online Data Catalog: Catalogue of stellar ultraviolet fluxes (TD1): A compilation of absolute stellar fluxes measured by the Sky Survey Telescope (S2/68) aboard the ESRO satellite TD-1. {\it Vizier Online Data Catalog} {\bf 2059} (1995). 

\bibitem{Mermilliod:1991} {Mermilliod}, J.~C. VizieR Online Data Catalog: Homogeneous Means in the UBV System (Mermilliod 1991). {\it VizieR Online Data Catalog} {\bf 2168} (2006). 

\bibitem{Hog:2000} {{H{\o}g}, E., {Fabricius}, C., {Makarov}, V.~V., {Urban}, S., {Corbin}, T., {Wycoff}, G., {Bastian}, U., {Schwekendiek}, P., {Wicenec}, A.} {The Tycho-2 catalogue of the 2.5 million brightest stars}. {\it Astronomy \& Astrophysics} {\bf 355}, {L27-L30} (2000). 

\bibitem{Droege:2006} {{Droege}, T.~F., {Richmond}, M.~W., {Sallman}, M.~P., {Creager}, R.~P.} {TASS Mark IV Photometric Survey of the Northern Sky}. {\it \pasp} {\bf 118}, {1666-1678} (2006). 

\bibitem{Cutri:2003} {Cutri}, R.~M. {\it et al}. {VizieR Online Data Catalog: 2MASS All-Sky Catalog of Point Sources (Cutri+ 2003)}. {\it VizieR Online Data Catalog} {\bf 2246} (2003). 
 
\bibitem{Skrutskie:2006} {{Skrutskie}, M.~F. {\it et al}.} {The Two Micron All Sky Survey (2MASS)}. {\it Astronomical Journal}, {\bf 131}, {1163-1183} (2006). 

\bibitem{Wright:2010} {{Wright}, E.~L., {\it et al}.} {The Wide-field Infrared Survey Explorer (WISE): Mission Description and Initial On-orbit Performance}. {\it Astronomical Journal} {\bf 140}, {1868-1881} (2010). 

\bibitem{Cutri:2012} {{Cutri}, R.~M., {\it et al.}} {VizieR Online Data Catalog: WISE All-Sky Data Release (Cutri+ 2012)}. {\it VizieR Online Data Catalog} {\bf 2311} (2012). 

\bibitem{Leeuwen:2007} {{van Leeuwen}, F.} {Validation of the new Hipparcos reduction}. {\it Astronomy \& Astrophysics} {\bf 474}, {653-664} (2007). 

\bibitem{Zacharias:2013} {{Zacharias}, N., {Finch}, C.~T., {Girard}, T.~M., {Bartlett}, J.~L., {Monet}, D.~G., {Zacharias}, M.~I.} {The Fourth US Naval Observatory CCD Astrograph Catalog (UCAC4)}. {\it Astronomical Journal} {\bf 145}, 44 (2013). 

\bibitem{Zacharias:2004} {{Zacharias}, N., {Monet}, D.~G., {Levine}, S.~E., {Urban}, S.~E., {Gaume}, R., {Wycoff}, G.~L.} {The Naval Observatory Merged Astrometric Dataset (NOMAD)}. In {\it American Astronomical Society Meeting Abstracts}, vol. 36 of the {\it Bulletin of the American Astronomical Society}, 1418 (2004). 

\bibitem{Kurucz:1992} {{Kurucz}, R.~L.} {Model Atmospheres for Population Synthesis}. In {{Barbuy}, B. and {Renzini}, A.} (eds.) {\it The Stellar Populations of Galaxies}, vol. 149 of {\it IAU Symposium}, 225 (1992). 

\bibitem{Schlegel:1998} {{Schlegel}, D.~J., {Finkbeiner}, D.~P., {Davis}, M.} {Maps of Dust Infrared Emission for Use in Estimation of Reddening and Cosmic Microwave Background Radiation Foregrounds}. {\it Astrophysical Journal} {\bf 500}, {525-553} (1998). 

\bibitem{Nelder:1965} {Nelder, J. A., Mead, R.} {A Simplex Method for Function Minimization}. {\it The Computer Journal} {\bf 7}, {308-313} (1965). 

\bibitem{Boue:2013} {{Bou{\'e}}, G., {Montalto}, M., {Boisse}, I., {Oshagh}, M., {Santos}, N.~C.} {New analytical expressions of the Rossiter-McLaughlin effect adapted to different observation techniques}. {\it \aap} {\bf 550}, A53 (2013). 

\bibitem{Gray:2008} {{Gray}, D.~F.} {\it The Observation and Analysis of Stellar Photospheres} (Cambridge, UK: Cambridge University Press, 2008). 

\bibitem{Collier:2010b} {{Collier Cameron}, A., {Bruce}, V.~A., {Miller}, G.~R.~M., {Triaud}, A.~H.~M.~J., {Queloz}, D.} {Line-profile tomography of exoplanet transits - I. The Doppler shadow of HD 189733b}. {\it \mnras} {\bf 403}, {151-158} (2010). 

\bibitem{Rossiter:1924} {{Rossiter}, R.~A.} {On the detection of an effect of rotation during eclipse in the velocity of the brigher component of beta Lyrae, and on the constancy of velocity of this system.} {\it Astrophysical Journal} {\bf 60}, {15-21} (1924). 

\bibitem{McLaughlin:1924} {{McLaughlin}, D.~B.} {Some results of a spectrographic study of the Algol system.} {\it Astrophysical Journal} {\bf 60}, {22-31} (1924). 

\bibitem{Bieryla:2015} {{Bieryla}, A. {\it et al}.} {KELT-7b: A Hot Jupiter Transiting a Bright V = 8.54 Rapidly Rotating F-star}. {\it Astronomical Journal} {\bf 150}, 12 (2015). 

\bibitem{Lomb:1976} {{Lomb}, N.~R.} {Least-squares frequency analysis of unequally spaced data}. {\it \apss} {\bf 39}, {447-462} (1976). 

\bibitem{Scargle:1982} {{Scargle}, J.~D.} {Studies in astronomical time series analysis. II - Statistical aspects of spectral analysis of unevenly spaced data}. {\it Astrophysical Journal} {\bf 263}, {835-853} (1982). 

\bibitem{Kunitomo:2011} {{Kunitomo}, M., {Ikoma}, M., {Sato}, B., {Katsuta}, Y., {Ida}, S.} {Planet Engulfment by \~{}1.5-3 M $_{sun}$ Red Giants}. {\it \apj} {\bf 737}, 66 (2011). 

\bibitem{Privitera:2016} {{Privitera}, G., {Meynet}, G., {Eggenberger}, P., {Vidotto}, A.~A., {Villaver}, E., {Bianda}, M.} {Star-planet interactions. II. Is planet engulfment the origin of fast rotating red giants?} {\it \aap} {\bf 593}, A128 (2016). 

\bibitem{Villaver:2007} {{Villaver}, E. and {Livio}, M.} {Can Planets Survive Stellar Evolution?} {\it \apj} {\bf 661}, {1192-1201} (2007). 

\end{thebibliography}
\end{document}